\documentclass[12pt]{article}
\usepackage{epsf}
\usepackage{cite}
\usepackage{amssymb}
\def\g{\gamma}
\def\a{\Gamma}

\def\pp{\pi^+}
\def\pn{\pi^-}
\def\po{\pi^0}
\def\gp{\g d\to\pp\g nn}
\def\gn{\g d\to\pn\g pp}
\def\go{\g d\to\po\g pn}
\def\beq{\begin{eqnarray}}
\def\eeq{\end{eqnarray}}

\begin{document}
\begin{titlepage}
\noindent
\begin{center}
{\Large \bf
 Search for narrow six-quark states in the reactions
\footnote
{Talk presented at the 5th Crystal Ball Meeting, Mainz, 19-21 March 2004}\\
$\g d\to \pi\g NN$}
\vskip 0.5cm

{\large
L.V.~Fil'kov
\footnote{E-mail: filkov@sci.lebedev.ru}  and
V.L.~Kashevarov
\footnote{E-mail: kashev@kph.uni-mainz.de}
}
\vskip 0.5cm

Lebedev Physical Institute, Leninsky Prospect 53, Moscow, Russia
\end{center}
\vskip 1cm

\begin{abstract}

We study the reactions $\g d\to\pi\g NN$
with the aim
to search for six-quark states, the decay of which into two nucleons is
forbidden by the Pauli exclusion principle. Such states predicted by a
variety of QCD inspired models and recent evidence from Proton Linear
Accelerator of INR (Moscow) strongly suggests the existence of such states
with the masses 1904, 1926, and 1942 MeV.
We propose an experiment at MAMI-B which will provide a unique opportunity to
observe such dibaryon states in mass region up to 2000 MeV and determine
their masses and quantum numbers.

\end{abstract}
\end{titlepage}

\newpage

\section{Introduction}
\label{sec:intro}

The possibility of the existence of multiquark states was predicted by QCD
inspired models \cite{jaf,muld}. These works initiated a lot of experimental
searches for six-quark states (dibaryons). Usually one looked for
dibaryons in the $NN$ channel (see for review ref. \cite{tat1}).
Such dibaryons have decay widths from a few up
to hundred MeV. Their relative contributions are small enough and the
background contribution is big and uncertain as a rule. All this often
leads to contradictory results.

In the present work we consider six-quark states,
a decay of which into two nucleons is forbidden by the Pauli exclusion
principle \cite{fil1,akh,fil2,alek,alek1}.
Such states satisfy the following condition:
\begin{equation}
(-1)^{T+S}P=+1
\end{equation}
where $T$ is the isospin, $S$ is the internal spin, and $P$ is the
dibaryon parity. In the $NN$ channel, these six-quark states would
correspond to the following forbidden states:
even singlets and odd triplets with the isotopic spin $T=0$ as well as
odd singlets and even triplets with \mbox{$T=1$}.
These six-quark states with the masses \mbox{$M < 2m_{N}+m_{\pi}$}
($m_N (m_{\pi}$) is the nucleon (pion) mass) can mainly decay
by emitting a photon. This is a new class of metastable six-quark states
with the decay widths \mbox{$< 1$keV}. Such states were called supernarrow
dibaryons (SND).
The experimental discovery of the SNDs would have important consequences
for particle and nuclear physics and astrophysics.

In the framework of the MIT bag model, Mulders et al. \cite{muld} calculated
the masses of different dibaryons, in particular, $NN$-decoupled dibaryons.
They predicted dibaryons $D(T=0;J^P =0^{-},1^{-},2^{-};M=2.11$ GeV) \ and
$D(1;1^{-};2.2$ GeV) corresponding to the forbidden states $^{13}P_J$
and $^{31}P_1$ in the $NN$ channel.
However, the dibaryon masses obtained exceed the pion production threshold.
Therefore, these dibaryons preferentially decay into the $\pi NN$ channel
and their decay widths are large than 1 MeV.

Using the chiral soliton model, Kopeliovich \cite{kop} predicted that
the masses of $D(T=1,J^P=1^+)$ and $D(0,2^+)$ SNDs
exceeded the two nucleon mass by 60 and 90 MeV, respectively.
These values are lower than the pion production threshold.

In the framework of the canonically quantized biskyrmion model,
Krupnovnickas {\em et al.} \cite{riska} obtained an indication on
possibility
of the existence of one dibaryon with J=T=0 and two dibaryons with J=T=1
with masses smaller than $2m_N+m_{\pi}$.

Unfortunately, all values obtained  for the dibaryon masses are model
dependent. Therefore, only an experiment could answer the question about
the existence of SNDs and determine their masses. In the following, we
summarize the experimental attempts to look for such SNDs so far.


In ref. \cite{konob,izv,ksf,yad,prc,conf1, conf2, epj}, the reaction
$pd\to p+pX_1$ and
$pd\to p+dX_2$ were studied with the aim of searching for SND.
The experiment was
carried out at the Proton Linear Accelerator of INR with 305 MeV
proton beam using the two-arm mass
spectrometer TAMS. As was shown in ref. \cite{yad,prc},
the nucleons and the deuteron from the decay of SND into $\gamma NN$
and $\gamma d$ have to be emitted into a narrow angle cone with respect to
the direction of the dibaryon motion. On the other hand, if a dibaryon
decays mainly into two nucleons, then the expected  angular cone of
emitted nucleons must be more than $50^{\circ}$. Therefore, a detection
of the scattered proton in coincidence with the proton (or the deuteron)
from the decay of the dibaryon at correlated angles allows the
suppression of the contribution of the background processes and increases
the relative contribution of a possible SND production.

Several software cuts have been applied to the mass spectra in these works.
In particular, the authors limited themselves by the consideration of an
interval of the proton energy from the decay of the $pX_1$ states, which
was determined by the kinematics of the SND decay into $\gamma NN$ channel.
Such a cut is very important as it provides a possibility to suppress the
contribution from the background reactions and random coincidences
essentially.

In the works \cite{conf1,conf2,epj}, CD$_2$ and $^{12}$C were
used as targets.
The scattered proton was detected in the left arm
of the spectrometer TAMS at the angle $\theta_L=70^{\circ}$. The second
charged particle (either $p$ or $d$) was detected in
the right arm by three telescopes located at $\theta_R=34^{\circ}$,
$36^{\circ}$, and $38^{\circ}$.

As a result,
three narrow peaks in missing mass spectra have been observed (Fig. 1a) at
$M_{pX_1}=1904\pm 2$, $1926\pm 2$, and $1942\pm 2$ MeV  with widths equal to
the experimental
resolution ($\sim 5$ MeV) and with numbers of standard deviations (SD) of
6.0, 7.0, and 6.3, respectively.
It should be noted that the dibaryon peaks at $M=1904$ and 1926 MeV had
been observed earlier by same authors in ref. \cite{prc,konob,izv,ksf,yad}
at somewhat different kinematical conditions. On the other hand, no noticeable
signal of the dibaryons has been observed in the missing mass $M_{dX_2}$
spectra of the reaction $pd\to p+dX_2$.
The analysis of the angular distributions of the protons from the decay of
$pX_1$ states and the suppression observed of the SND decay into $\gamma d$
showed that the peaks found can be explained as a manifestation of the
isovector SNDs, the decay of which into two nucleons is forbidden
by the Pauli exclusion principle.

\begin{figure}
\epsfxsize=14cm
\epsfysize=16cm
\centerline{
\epsffile{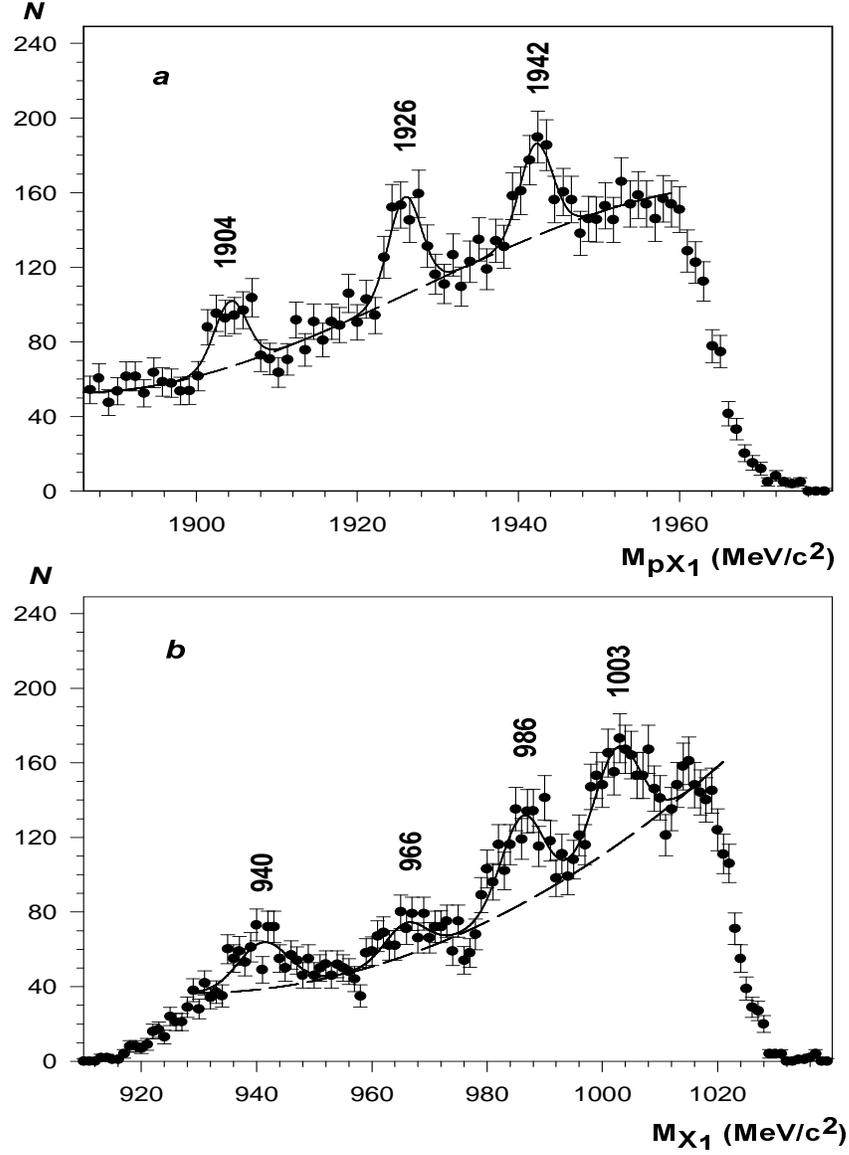}}
\caption{
The missing mass $M_{pX_1}$ (a) and
$M_{X_1}$ (b) spectra of the reaction $pd\to p+pX_1$
for the sum of angles of
$\theta_R=34^{\circ}$ and $\theta_R=36^{\circ}$. The dashed and solid curves
are results of interpolation by polynomials
(for the background) and Gaussian (for the peaks), respectively.}
\end{figure}

An additional information about the nature of the observed states
was obtained by studying the missing mass $M_{X_1}$ spectra of the
reaction $pd\to p+pX_1$.
If the state found is a dibaryon decaying mainly into two nucleons then
$X_1$ is a neutron and the mass $M_{X_1}$ is equal to the neutron mass
$m_n$. If the value of $M_{X_1}$, obtained from the experiment, differs
essentially from $m_n$ then $X_1=\gamma+n$ and we have the additional
indication that the observed dibaryon is the SND.

The simulation of the missing mass $M_{X_1}$ spectra of the reaction
$pd\to ppX_1$ has been performed \cite{conf1,conf2,epj} assuming that
the SND decays as $D\to\g+\, ^{31}S_0\to \g pn$ through two nucleon singlet
state $^{31}S_0$ \cite{fil2,prc,epj}. As a result, three narrow peaks at
$M_{X_1}=965$, 987, and 1003 MeV have been predicted. These peaks
correspond to the decay of the isovector SNDs with masses 1904, 1926, and
1942 MeV, respectively.

In the experimental missing mass $M_{X_1}$ spectrum besides the peak
at the neutron mass caused by the process $pd\to p+pn$,
three peaks have been observed at $966\pm 2$,
$986\pm 2$, and $1003\pm 2$ MeV \cite{conf1,conf2,epj}.
These values of $M_{X_1}$ coincide with
the ones obtained by the simulation and essentially differ from
the value of the neutron mass (939.6 MeV). Hence, for all states under
study, we have $X_1=\gamma+n$ in support of the statement that the
dibaryons found are SNDs.

On the other hand, the peak at $M_{X_1}=1003\pm 2$ MeV corresponds to
the peak found in ref. \cite{tat2} and was attributed to an exotic
baryon state $N^*$ below the $\pi N$ threshold. In that work the authors
investigated the reaction $pp\to\pi^+X$ and have found
altogether three such states with masses 1004, 1044, and 1094 MeV.
Therefore, if the exotic baryons with anomalously small masses really
exist, the observed peaks at 966, 986, and 1003 MeV might be a manifestation
of such states.
The existence of such exotic states, if
proved to be true, will fundamentally change our understanding of the
quark structure of hadrons \cite{bald,walch}.
In ref. \cite{azim} these states were considered as possible candidates
for pentaquark baryons consisted of $u$ and $d$ quarks.

However, the experiments on single nucleon have
not observed any significant structure \cite{lvov,jiang,kohl}. Therefore,
the question about a nature of peaks observed in \cite{epj,tat2} is
open at present.


In ref. \cite{khr} dibaryons with exotic quantum numbers were searched for
in the process $pp\to pp\gamma\gamma$. The experiment was performed with
a proton beam from the JINR Phasotron at an energy of about 216 MeV. The
energy spectrum of the photons emitted at $90^{\circ}$ was measured.
As a result, two peaks have been observed in this spectrum. This behavior
of the photon energy spectrum was interpreted as a signature of the exotic
dibaryon resonance $d_1$ with a mass of about 1956 MeV and possible isospin
$T=2$. However, the result obtained in ref. \cite{str} makes
the possibility of production of dibaryons with $T=2$ in this reaction
questionable. So, additional careful studies of the
reaction $pp\to pp\gamma\gamma$ are needed to understand more correctly
the nature of the observed state.

On the other hand, an analysis \cite{cal} of the Uppsala proton-proton
bremsstrahlung data looking for the presence of a dibaryon in the mass range
from 1900 to 1960 MeV only gave the upper limits of 10 and 3 nb for
the dibaryon production cross section at proton beam energies of 200 and
310 MeV, respectively.
This result agrees with the estimates of the cross
section obtained at the conditions of this experiment in the framework
of the dibaryon production model suggested in ref. \cite{prc} and does
not contradict the data of ref. \cite{khr}.

Gerasimov \cite{ger1,ger2} suggested that a double radiative capture on
the pionic deuterium $((\pi^-d)_{atom}\to nn\gamma\gamma)$ is a candidate
for the further investigation of the narrow dibaryons existence. The
dibaryon in this reaction can be produced via the radiative capture
$\pi^-d\to \gamma d_1$ and then
it decays to $\gamma nn$. Such an experiment has been carried out at the
TRIUMF cyclotron using  the beam of negative pions with a central
momentum of 81.5 MeV/c \cite{triumf}. No evidence
for narrow dibaryons has been found in this work and a branching ratio
upper limit, $BR<6.7\times 10^{-6}$ $(90\%\;C.L.)$, for narrow $d_1$
production in the mass region from 1920 to 1980 MeV was obtained. This
upper limit is several order of magnitude below the yield estimate of
Gerasimov $(\sim 0.5\%)$ obtained in a simple model \cite{ger1}.
However, if $d_1$ is SND, this model does not really take into account
a very strong overlap of the baryons which is a necessary condition of
the SND production \cite{fil2,prc,epj}.

Let us make a crude estimation of the branching ratio of SND $D(1,1^{\pm})$
production by a radiative capture on the pionic deuterium,
$BR((\pi^-d)_{atom}\to \gamma D(1,1^{\pm})$, assuming that two neutrons in
the reaction $(\pi^-d)_{atom}\to nn$ mainly are in $^{31}S_0$ state and
using the model suggested in \cite{fil2,prc,epj}. Then for the
reaction $(\pi^-d)_{atom}\to nn\to \gamma D(1,1^{\pm})$ we have
$$
BR((\pi^-d)_{atom}\to \gamma D(1,1^{\pm}))\sim BR((\pi_d)_{atom}\to nn)
\alpha\eta_S
$$
where $BR((\pi^-d)_{atom})\to nn\approx 0.74$ \cite{high}, $\alpha$ is
the fine-structure constant, $\eta_S$ is the probability of the full
overlap of two neutrons in $^{31}S_0$ state. Using the experimental
value of the SND production cross section in the process
$pd\to p+pX_1$ \cite{epj,tamii}, one has shown \cite{mass} that the
probability of such an overlap of two nucleons in the deuteron
($\eta$) is equal to $\sim 10^{-4}$. Supposing that also
$\eta_S\approx 10^{-4}$ we obtain
$$
BR((\pi^-d)_{atom}\to \gamma D(1,1^{\pm}))\sim 0.74/137\times 10^{-4}=
5.4\times 10^{-7}.
$$
This value is one order of magnitude below the upper limit obtained
in \cite{triumf}.

The reactions $pd\to pdX$ and $pd\to ppX$ have
been investigated also by Tamii {\em et al.} \cite{tamii}
at the Research Center for Nuclear Physics
at the proton energy 295 MeV in the mass region of 1896--1914
MeV. They did not observe any narrow structure in this mass region
and obtained the upper limit of the production cross section of a
NN-decoupled dibaryon is equal to $\sim 2$ $\mu$b/sr if the dibaryon
decay width $\Gamma_D<< 1$ MeV. And if $\Gamma_D=3$ MeV, the
upper limit will be about $3.5\mu$b/sr. These limits are
smaller than the value of the cross section of $8\pm 4\mu$b/sr
declared in ref. \cite{prc}.

However, the latter value was overestimated that was caused by
not taking into account
angle fluctuations related to a beam position displacement on
the CD$_2$ target during the run.
The beam displacement during a run decreased the efficiency for the elastic
scattering and practically did not do that for the dibaryon formation
reaction because of a wider cone of outgoing particles for the latter.
As was shown in the next experimental runs, the real
value of the cross sections of the production of the SND with
the mass 1904 MeV must be smaller by 2--3 times than
that was estimated in ref. \cite{prc}.

On the other hand, simulation shows that the energy distribution of the
protons from the decay of the SND with the mass of 1904 MeV is
rather narrow with the maximum at $\sim 74$ MeV. This distribution
occupies the energy region of 60--90MeV. The experiments \cite{prc,epj}
confirmed the result of this simulation. However, in ref. \cite{tamii} the
authors considered the region 74--130 MeV. Therefore, they could detect
only a small part of the SND contribution. Moreover, they used a very
large angular acceptance of the spectrometer which detected these protons,
while the protons under consideration have to fly in very narrow angle
cone. As a result, the effect-to-background ratio in this experiment was
more than 10 times less than in ref. \cite{prc,epj}. Very big errors
and absence of proper proton energy and angular cuts
in ref. \cite{tamii} did not allow authors to observe any structure
in the $pX_1$ mass spectrum.

It is worth noting that the reaction $pd\to NX$ was investigated in
other works, too (see for example \cite{set}). However, in contrast to
the ref. \cite{konob,izv,ksf,yad,prc,conf1,conf2,epj},
the authors of these works did not study
either the correlation between the parameters of the scattered proton and
the second detected particle
or the emission of the photon from the dibaryon decay.
Therefore, in these works the relative contribution of the dibaryons under
consideration was small, which hampered their observation.

However, in order to argue more convincingly that the states found are
really SNDs, an additional experimental investigation of the dibaryon
production is needed.

On this point,
the search for the SNDs in processes of the pion photoproduction from
the deuteron is of great interest. Besides a decision of the question
about an existence of the SNDs, an investigation of this process
will allow the quantum numbers of the SNDs to be determined.
And the experiment with the linear polarized photons will give
possibility to determine other quantum numbers of the SNDs \cite{alek1}.

In ref. \cite{pi0} narrow dibaryon resonances were searched for in the
reaction $\g d\to\pi^0X$ in the photon energy region 140--300 MeV.
No significant structure has been observed. Upper limits for the production
of narrow dibaryons in the range 2--5 $\mu$b averaged over the 0.8 MeV
resolution has been observed. However, the expected values of the SND
production cross section are essentially less. The estimation of
the total cross section of the SND production in this process in the
photon energy region 200--300 MeV (see sec. 3) gives 0.01--0.03 $\mu$b
for the dibaryon $D(1,1^+)$ with the mass 1904 MeV and it is less for
the vector SND and the higher masses.

Essentially bigger cross sections are expected for the SND production
in the processes of the charged pion photoproduction.

We propose to search for the SNDs by studying the reactions $\gp$,
$\gn$, and $\go$ in the photon energy region 300--800 MeV at MAMI.
The mass of the SNDs will be reconstructed by the measurement of the
photon and two nucleons. The detection of the pions will permit
the background to be suppressed additionally.
Using of a deuteron target allows
avoiding uncertainties taken place in ref. \cite{prc,conf1,conf2,epj,tamii}
where CD$_2$ target was
used. Furthermore, use of the photon beam with the energies of 300--800
MeV and the Crystal Ball spectrometer and TAPS will give a possibility
to suppress essentially background and to investigate a wide mass spectrum
and, particular, check the possibility of existence of the SNDs at
$M_{pX_1}=1956$ \cite{khr} and 1982 MeV. The latter one was predicted in
\cite{epj,mass}.
This experiment can also help to understand nature of the peaks
in the $M_{X_1}$ mass spectra and clarify a possibility of existence
of exotic baryons with small masses.

\section{The cross sections of the supernarrow dibaryon\\ production
in the reactions $\lowercase{\g d}\to \lowercase{\pi}D$ \protect\\}
\label{sec:2}

We will consider the following SNDs:
$D(T=1,J^P=1^+,S=1)$ and $D(1,1^-,0)$.

It is worth noting that the  state $(T=1, J^P=1^-)$ corresponds
to the states $^{31}P_1$ and $^{33}P_1$ in the NN channel.
The former is forbidden and
the latter is allowed for a two-nucleon state. In our work we will study
the dibaryon $D(1,1^-,0)$, a decay of which into two nucleons is forbidden
by the Pauli principle (i.e. $^{31}P_1$ state).

In the process $\g d\to pD$, SNDs can be produced
only if the  nucleons in the deuteron overlap sufficiently, such that a
6-quark state with deuteron quantum numbers can be formed. In this case, an
interaction of a photon or a meson with this state can change its
quantum numbers so that a metastable state is formed.
Therefore, the probability of the production of such dibaryons is
proportional to the probability $\eta$ of the 6-quark state existing
in the deuteron.

The magnitude of $\eta$ can be estimated from the
deuteron form factor at large $Q^2$ (see for example \cite{bur}). However,
the values obtained depend strongly on the model of the form factor of
the 6-quark state over a broad region of $Q^2$. Another way to estimate
this parameter is to use the discrepancy between the theoretical and
experimental values of the deuteron magnetic moment \cite{kim,kon2}.
This method is free from the restrictions quoted above and gives
$\eta\le 0.03$ \cite{kon2}. In ref. \cite{mass} the magnitude of $\eta$
was estimated by an analysis of the mass formula for SNDs using the
experimental value of the cross section of the SND production in
the process $pd\to p+pX_1$ obtained in \cite{epj,tamii}. As a result,
$\eta\approx 10^{-4}$ has been obtained.

\begin{figure}[h]
\centerline{
\epsfxsize=6cm  
\epsfysize=3cm  
\epsffile{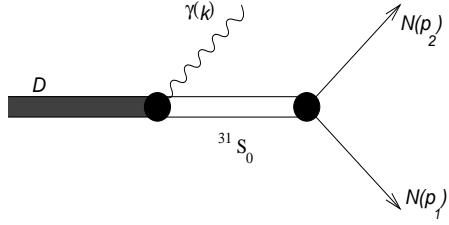}}
\caption{The diagram of the SND decay into $\g NN$}\label{dec}
\end{figure}

\begin{figure}[h]
\centerline{
\epsfxsize=10cm
\epsfysize=10cm
\epsffile{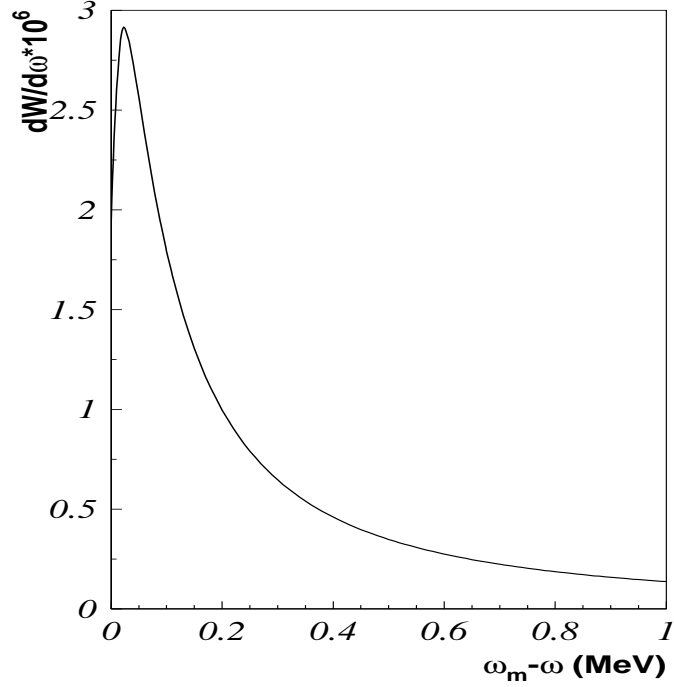}}
\caption{The distribution of the decay probability of the isovector
SND over $\omega_m-\omega$ at $M=1904$ MeV.}\label{width}
\end{figure}

Since the energy of nucleons, produced in the decay of the SNDs under
study with $M<2m_N +m_{\pi}$, is small, it may be expected that the main
contribution to a two nucleon system should come from the
$^{31}S_0$ (virtual singlet) state (Fig. \ref{dec}).
The results of calculations of the
decay widths of the dibaryons into $\g NN$ on the
basis of such assumptions at $\eta=0.01$ are listed in Table \ref{table1}.
If $\eta=10^{-4}$, the values of the decay widths would be smaller by
$\sim 10^2$ times.
\begin{table}
\centering
\caption{ Decay widths of the dibaryons $D(1,1^+,1)$ and $D(1,1^-,0)$
at various dibaryon masses $M$. $\a_{t}\approx\a_{\g NN}$
\label{table1}}
\begin{tabular}{|c|c|c|c|c|c|c|c|}\hline
$M$(GeV)     & 1.90 & 1.91 & 1.93 & 1.96 & 1.98 & 2.00 & 2.013 \\ \hline
$\a_t(1,1^+)$& 0.51 & 1.57 & 6.7  & 25.6 & 48   & 81   & 109   \\
(eV)         &      &      &      &      &      &      &       \\ \hline
$\a_t(1,1^-)$& 0.13 & 0.39 & 1.67 & 6.4  & 12   & 20   & 27    \\
(eV)         &      &      &      &      &      &      &       \\ \hline
\end{tabular}
\end{table}
As a result of the SND decay through $^{31}S_0$ in
the intermediate state, the probability distribution of such a decay
over the emitted photon energy $\omega$ should be characterized by a
narrow peak at the photon energy close to the maximum value
$\omega_m=(M^2-4m^2_N)/2M$ (Fig.~\ref{width}).
Note that the interval of the photon energy from $\omega_m$ to
$\omega_m-1$ MeV contains about 75\% of the contribution to the width
of the decay $D(1,1^{\pm})\to \gamma NN$. This leads to a very small
relative energy of the nucleons from the SND decay ($\lesssim 1$ MeV)
and these nucleons will be emitted into a narrow angle cone with respect
to the direction of the SND motion.

\begin{figure}[h]
\centerline{
\epsfxsize=17cm   
\epsfysize=10cm   
\epsffile{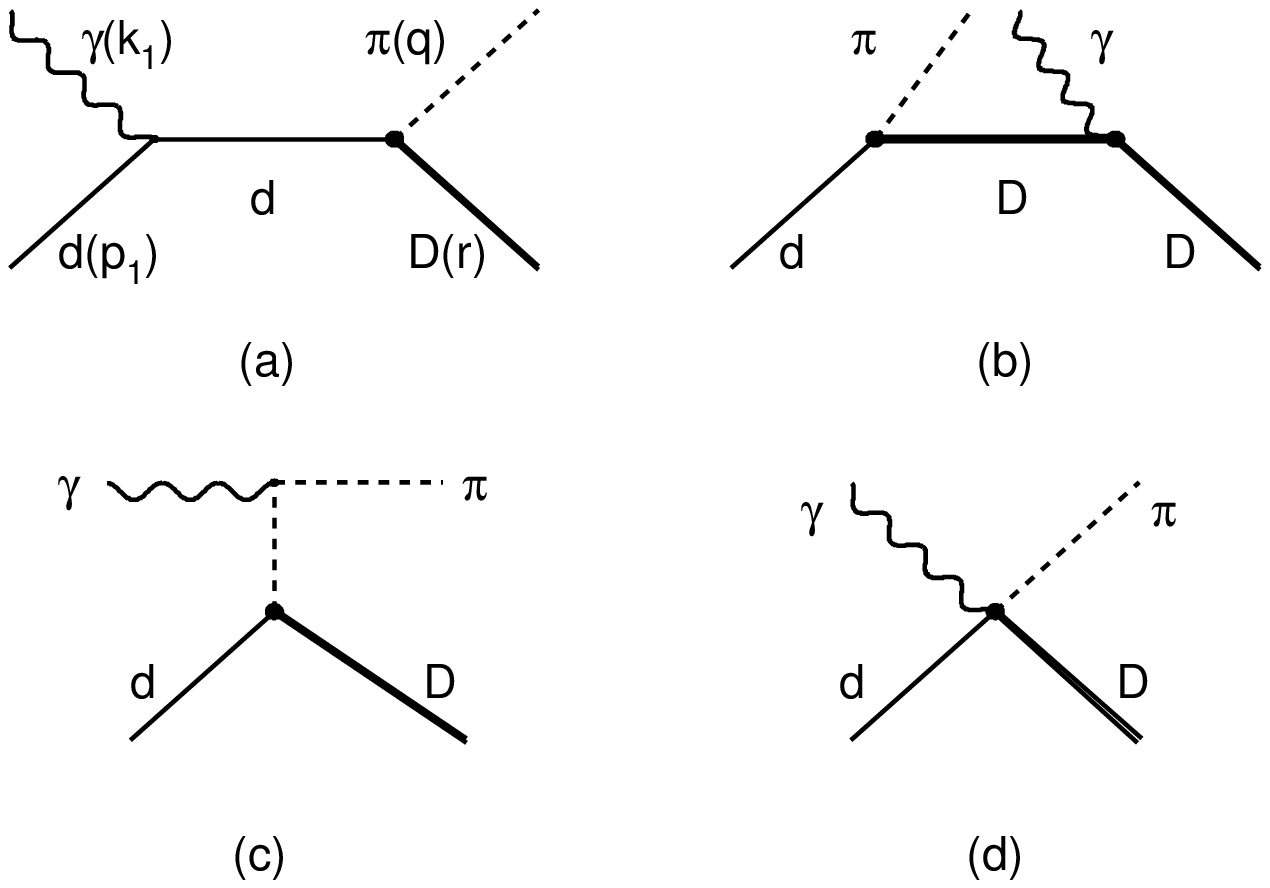}}     
\caption{The diagrams of the SND production in the process
$\g d\to \pi D$}\label{diag}
\end{figure}

Let us calculate the cross section of the SND photoproduction.
The gauge invariant amplitude of photoproduction of the SND
dibaryon may be obtained with help of diagrams in Fig.~\ref{diag}.

Such a dibaryon could be
produced only if a pion is emitted from the 6-quark state of the
deuteron. Therefore the vertexes of $d\to \pi D$ are written as
\begin{equation}
\a_{d\to\pi D(1,1^{-},0)}=\frac{g_1}{M} \sqrt{\eta}
\Phi_{\mu \nu} G^{\mu \nu} ,
\end{equation}
\begin{equation}
\a_{d\to\pi D(1,1^+,1)}=\frac{g_2}{M}\sqrt{\eta}\varepsilon_{\mu\nu
\lambda\sigma}\Phi^{\mu\nu}G^{\lambda\sigma},
\end{equation}
where $\Phi_{\mu \nu}=r_{\mu}w_{\nu}-w_{\mu}r_{\nu}$,
$G_{\mu \nu}=p_{1\mu}v_{\nu}-v_{\mu}p_{1\nu}$, $w$ and $v$ are 4-vectors
of the dibaryon and deuteron polarization, $r$ and $p_1$ are the dibaryon
and deuteron 4-momenta.

The following matrix elements correspond to the diagrams in
Fig. \ref{diag}a,b,c,d for the SND $D(1,1^-,0)$
\beq\label{tmin}
T_a&=&-e\sqrt{\eta}\frac{g_1}{M}\frac{(\epsilon (2p_1+k_1))}{(k_1p_1)}
\left\{(vw)(rP)-(vr)(wP)\right\}\nonumber\\
T_b&=&e\sqrt{\eta}\frac{g_1}{M}\frac{(\epsilon (2r-k_1))}{(k_1r)}\left\{
(vw)(p_1Q)-(vQ)(p_1w)\right\}
\nonumber\\
T_c(1^-)&=&e\sqrt{\eta}\frac{g_1}{M}\frac{(\epsilon (2q-k_1))}{(k_1q)}
\left\{(vw)(p_1r)-(vr)(p_1w)\right\} \\
T_d(1^-)&=&T_{d1}(1^-)+\alpha T_{d2}(1^-)\nonumber\\
T_{d1}(1^-)&=&2e\sqrt{\eta}\frac{g_1}{M}\left\{(vw)(\epsilon r)-
(vr)(\epsilon w)
\right\}
\nonumber\\
T_{d2}(1^-)&=&2e\sqrt{\eta}\frac{g_1}{M}\left\{(vw)(\epsilon p_1)-(\epsilon v)
(p_1w)\right\}
\nonumber
\eeq
where $P=(p_1+k_1)$, $Q=(r-k_1)$; $k_1$ and $q$ are the 4- momenta of the
photon and the pion, respectively; $\epsilon$ is the 4-vector of the photon
polarization. The expression for the matrix element
$T_d$ is written in the form which ensures the gauge invariance of the
amplitude of the dibaryon photoproduction. The coefficient $\alpha$ is
equal to 0, 1, 2 for the production of the $\pi^+$, $\pi^0$, $\pi^-$
mesons, respectively.

For the SND $D(1,1^+,1)$ we have
\beq\label{tpl}
T_a(1^+)&=&2e\sqrt{\eta}\frac{g_2}{M}\frac{(\epsilon(2p_1+k_1))}{(k_1p_1)}
\varepsilon_{\mu\nu\lambda\sigma}v^{\mu}w^{\nu}P^{\lambda}r^{\sigma}
\nonumber\\
T_b(1^+)&=&-2e\sqrt{\eta}\frac{g_2}{M}\frac{(\epsilon(2r-k_1))}{(k_1r)}
\varepsilon_{\mu\nu\lambda\sigma}v^{\mu}w^{\nu}p_1^{\lambda}Q^{\sigma}
\nonumber\\
T_c(1^+)&=&-2e\sqrt{\eta}\frac{g_2}{M}\frac{(\epsilon(2q-k1))}{(k_1q)}
\varepsilon_{\mu\nu\lambda\sigma}v^{\mu}w^{\nu}p_1^{\lambda}r^{\sigma}\\
T_d(1^+)&=&T_{d1}(1^+)+\alpha T_{d2}(1^+) \nonumber\\
T_{d1}(1^+)&=&-4e\sqrt{\eta}\frac{g_2}{M}
\varepsilon_{\mu\nu\lambda\sigma}v^{\mu}w^{\nu}\epsilon^{\lambda}r^{\sigma}
\nonumber\\
T_{d2}(1^+)&=&-4e\sqrt{\eta}\frac{g_2}{M}
\varepsilon_{\mu\nu\lambda\sigma}v^{\mu}w^{\nu}p_1^{\lambda}
\epsilon^{\sigma}
\nonumber
\eeq

The matrix elements (\ref{tmin}) and(\ref{tpl}) are connected with the
amplitudes of the SND photoproduction in different channels as:
\beq\label{chan}
T(\g d\to\pi^0 D)&=&T_a+T_b+T_{d1}+T_{d2} \nonumber\\
T(\g d\to \pi^+D)&=&\sqrt{2}\left(T_a+T_c+T_{d1}\right) \\
T(\g d\to\pi^-D)&=&-\sqrt{2}\left(T_a+2T_b-T_c+T_{d1}+2T_{d2}\right)
\nonumber
\eeq

Let us calculate the cross section of the dibaryon $D(1,1^-,0)$
photoproduction in the reactions with formation of $\pi^+$ meson. We will use
the calibration $\epsilon_0 =0$. Then the amplitude $T_a$ is equal to zero in
lab. system. As result of the calculation we have (in lab. system):
\beq\label{dmin}
&&\frac{d\sigma_{\g d\to\pi^+D(1^-)}}{d\Omega}
=\frac{1}{3}\left(\frac{e^2}{4\pi}\right)\left(\frac{g_1^2}
{4\pi}\right)\eta\frac{q_1^2}{m_dM^2\nu J}\left\{\frac{1}{2m_d^2}
[(M^2+m_d^2-t)^2-4m_d^2M^2]\right.+ \nonumber \\
&& \left. q_1^2(1-\cos^2\theta)\left[1+2\frac{(M^2+m_d^2-t)^2+2m_d^2M^2}
{(\mu^2-t)^2}-4\frac{M^2+m_d^2-t}{\mu^2-t}\right]\right\}
\eeq
where
$$
s=2m_d\nu+m_d^2, \quad t=\mu^2-2\nu (q_0-q_1\cos\theta_{\pi}), \quad
J=q_1(m_d+\nu)-q_0\nu \cos\theta_{\pi},
$$
$m_d$ is the deuteron mass, $\nu$ is the incident photon energy, $q_0(q_1)$
is the $\pi$ meson energy (momentum). The pion energy $q_0$ is connected with
the pion emission angle $\theta_{\pi}$ in the following way:
\begin{equation}
q_0=\frac{1}{c_1}\left[(m_d+\nu)c_2 +\nu\cos\theta_{\pi}
\sqrt{c_2^2-2\mu^2c_1}\right]
\end{equation}
where
$$
c_1=2[(m_d+\nu)^2-\nu^2\cos^2\theta_{\pi}], \qquad c_2=s+\mu^2-M^2 .
$$

The calculation of the cross section of the $D(1,1^+,1)$ SND photoproduction
has resulted in
\beq
&&\frac{d\sigma_{\g d\to\pi^+D(1^+)}}{d\Omega}
=\frac{8}{3}\left(\frac{e^2}{4\pi}\right)\left(\frac{g_2^2}
{4\pi}\right)\eta\frac{q_1^2}{m_dM^2\nu J}\left\{2M^2\right.+ \nonumber \\
&&\left. q_1^2(1-\cos^2\theta)\left[1-2\frac{m^2_d+M^2-t}{\mu^2-t}+
\frac{(m_d^2+M^-t)^2-4m_d^2M^2}{(\mu^2-t)^2}\right]\right\}
\eeq

For the $D(1,1^-,0)$ and $D(1,1^+,1)$ photoproduction in the process
$\gamma d\to \pi^- +D$ we have obtained:
\beq
&&\frac{d\sigma_{\g d\to\pi^-D(1^-)}}{d\Omega}
=\frac{1}{3}\left(\frac{e^2}{4\pi}\right)\left(\frac{g_1^2}
{4\pi}\right)\eta\frac{q_1^2}{m_dM^2\nu J}\left\{\frac12(M^2+m_d^2-t)
\left(\frac1{m_d^2} +\frac4{M^2}\right)\right.- \nonumber \\
&&\left. (M^2+4m_d^2)+q_1^2(1-\cos^2\theta)\left[1+\frac{A_1}{(M^2-u)^2}+
\frac{A_2}{(M^2-u)}+\right.\right. \\
&&\left.\left.\frac{A_3}{(M^2-u)(\mu^2-t)}+\frac{A_4}{(\mu^2-t)}+\frac{A_5}
{(\mu^2-t)^2}\right]\right\}, \nonumber
\eeq
\beq
&&\frac{d\sigma_{\g d\to\pi^-D(1^+)}}{d\Omega}
=\frac{8}{3}\left(\frac{e^2}{4\pi}\right)\left(\frac{g_2^2}
{4\pi}\right)\eta\frac{q_1^2}{m_dM^2\nu J}\left\{2\left[M^2+4m_d^2-4m_dr_0
\right]\right.+ \nonumber \\
&& \left. q_1^2(1-\cos^2\theta)\left[1+16\frac{q_1^2m_d^2}{(M^2-u)^2}+
4\frac{r_1^2m_d^2}{(\mu^2-t)^2}-16\frac{m_d^2(q_0r_0-s+M^2+\mu^2)}
{(M^2-u)(\mu^2-t)}- \right.\right.\\
&&\left.\left. 8m_d(r_0-2m_d)\left(\frac2{(M^2-u)}+\frac1{(\mu^2-t)}\right)
\right]\right\}, \nonumber
\eeq
where
$$
u=M^2+m_d^2+\mu^2-s-t, \quad r_0=m_d+\nu-q_0, \quad r_1=\sqrt{r_0^2-M^2},
$$
\beq
A_1&=&4\left[(m_d^2-\mu^2+u)^2+4m_d^2u+\frac{(M^2+m_d^2-t)}{M^2}\left(
(M^2+u)(m_d^2-\mu^2+u)\right.\right.-\nonumber \\
  && \left.\left.u(M^2+m_d^2-t)\right)\right]\nonumber \\
A_2&=&4\left[2(3m_d^2-\mu^2+u)-\frac1{M^2}(s-m_d^2)(M^2+m_d^2-t)\right]
\nonumber \\
A_3&=&8\left[(md2-\mu^2+u)(M^2+m_d^2-t)+m_d^2(M^2+u)\right]\\
A_4&=&4(M^2+3m_d^2-t)\nonumber \\
A_5&=&2\left[(m_d^2+\mu^2-t)^2+2m_d^2M^2\right].\nonumber
\eeq

The calculation of the $D(1,1^-,0)$ and $D(1,1^+,1)$ production in the process
$\gamma d\to\pi^0+D$ has given:
\beq
&&\frac{d\sigma_{\g d\to\pi^0D(1^-)}}{d\Omega}
=\frac{1}{6}\left(\frac{e^2}{4\pi}\right)\left(\frac{g_1^2}
{4\pi}\right)\eta\frac{q_1^2}{m_dM^2\nu J}\left\{\frac12(M^2+m_d^2-t)
\left(\frac1{m_d^2} +\frac1{M^2}\right)\right.- \nonumber \\
&&\left.2\left(M^2+m_d^2\right)+q_1^2(1-\cos^2\theta)\left[1+
\frac{A_1}{4(M^2-u)^2}+\frac{A_6}{(M^2-u)}\right]\right\}
\eeq
\beq
&&\frac{d\sigma_{\g d\to\pi^0D(1^+)}}{d\Omega}
=\frac{4}{3}\left(\frac{e^2}{4\pi}\right)\left(\frac{g_2^2}
{4\pi}\right)\eta\frac{q_1^2}{m_dM^2\nu J}\left\{\left(M^2+m_d^2+t
\right)\right.+ \nonumber \\
&&\left.q_1^2(1-\cos^2\theta)\left[1+\frac{4q_1^2m^2_d}{(M^2-u)^2}+
\frac{2(M^2-m^2_d-t)}{(M^2-u)}\right]\right\},
\eeq
and
$$
A_6=4(2m^2_d-\mu^2+u)-\frac{(M^2+m^2_d-t)}{M^2}(s-m^2_d).
$$

The constants $g^2_1/4\pi$, $g^2_2/4\pi$, and $\eta$ are unknown.
However, the product of these coupling constants and $\eta$ can be
estimated from the results of works \cite{epj,tamii} where the SNDs
were searched for in the process $pd\to p pX_1$. As a result we have
\begin{equation}
\eta\frac{g_1^2}{4\pi}=0.7\times 10^{-3}, \qquad
\eta\frac{g_2^2}{4\pi}=1.5\times 10^{-3}.
\end{equation}

\section{ Kinematics of SND Photoproduction}

We will consider SND with the masses $M$ in the region 1900--2000 MeV.

\begin{figure}[ht]
\begin{minipage}{0.45\textwidth}  
\epsfxsize=\textwidth
\epsffile{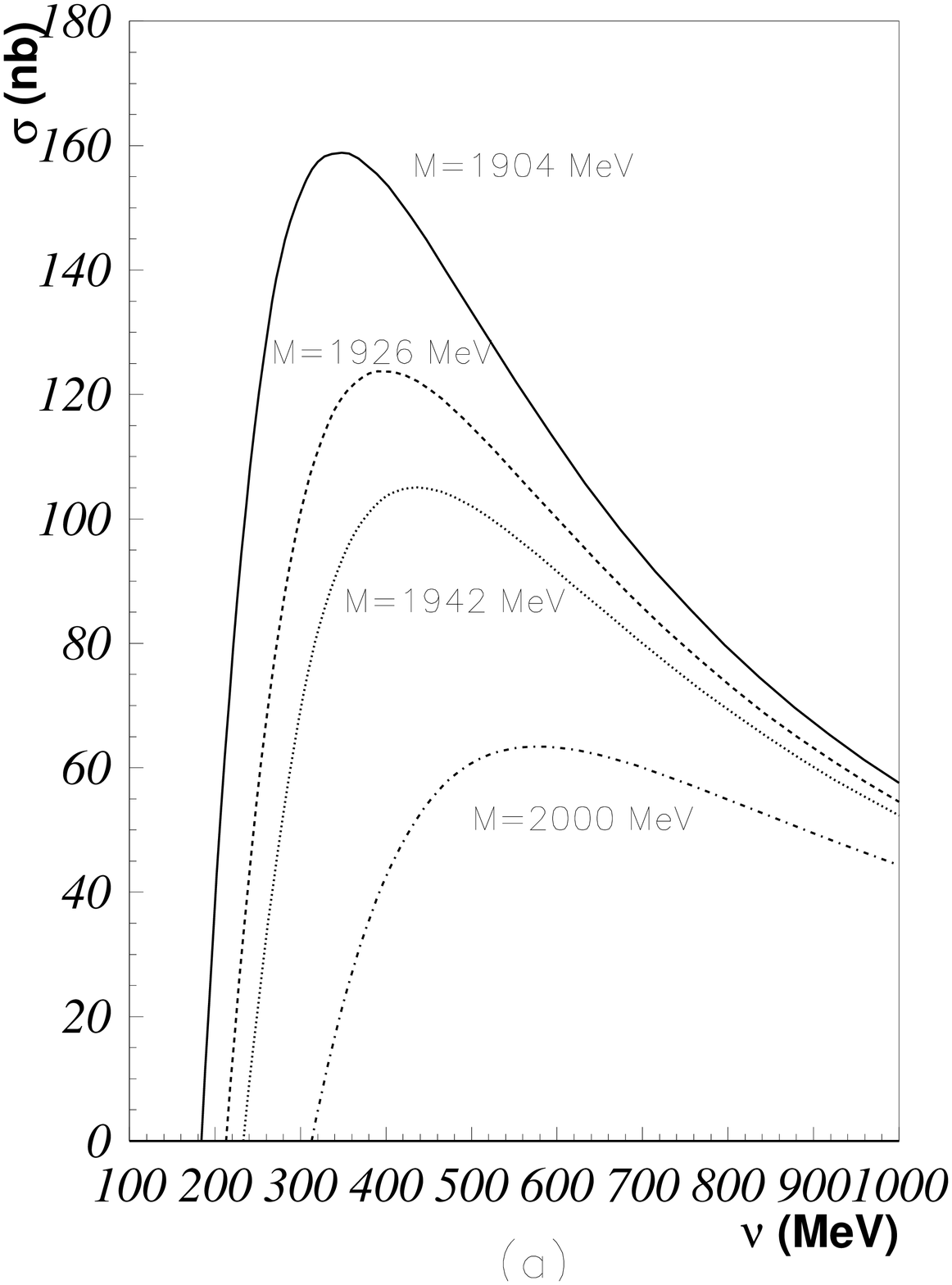}
\end{minipage}
\begin{minipage}{0.45\textwidth}
\epsfxsize=\textwidth
\epsffile{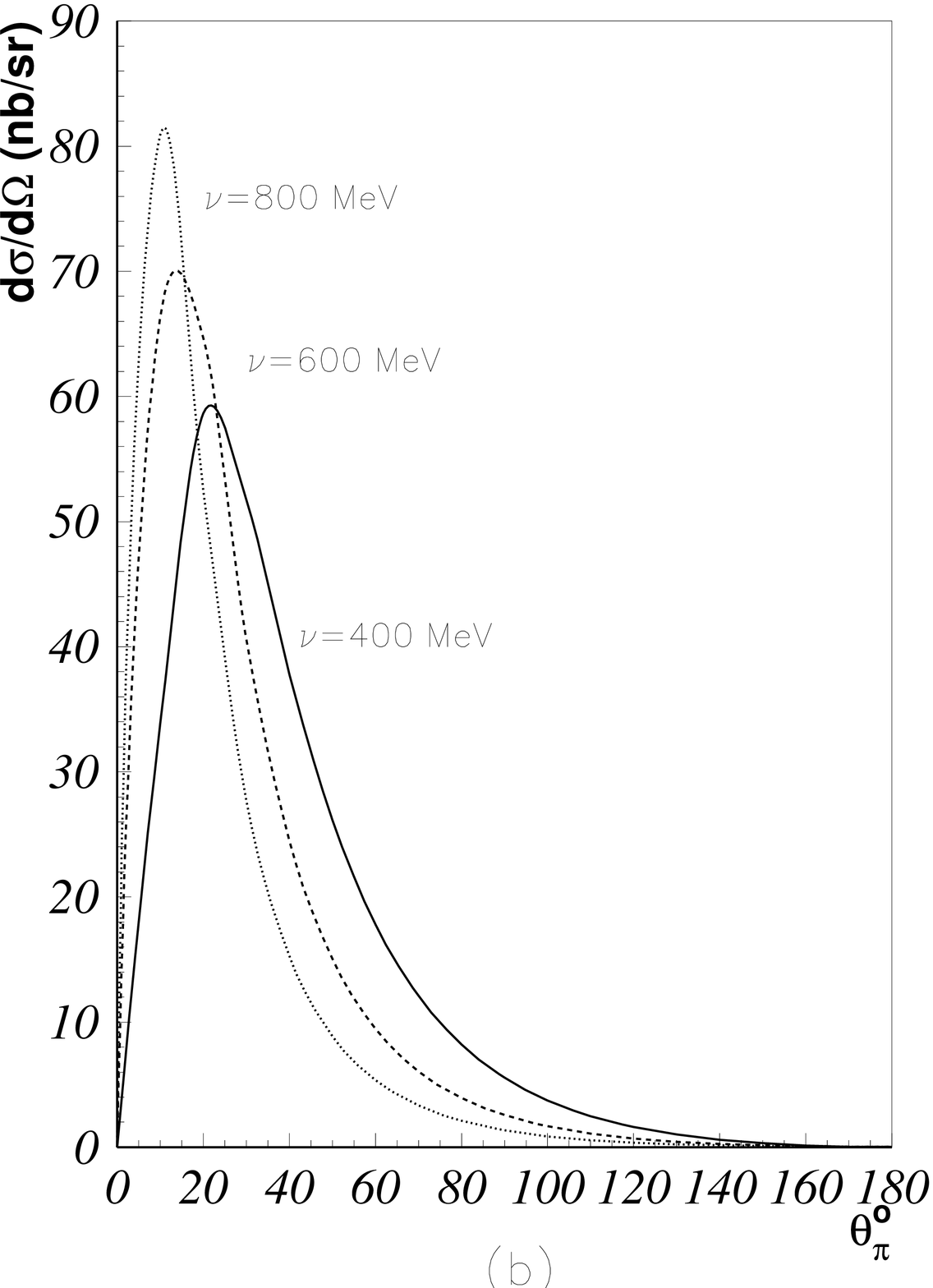}
\end{minipage}
\begin{minipage}{0.45\textwidth}
\epsfxsize=\textwidth
\epsffile{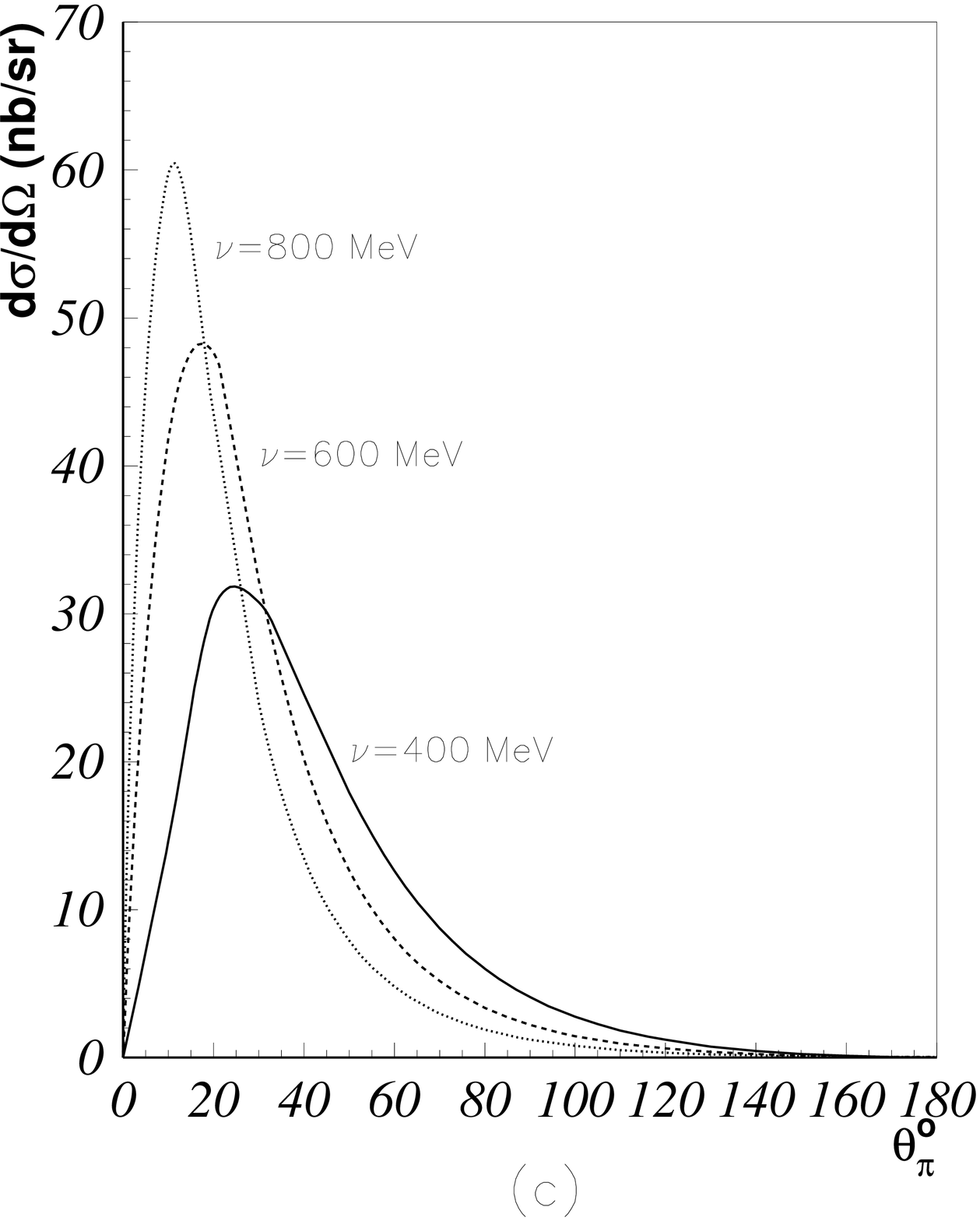}
\end{minipage}
\begin{minipage}{0.45\textwidth}
\epsfxsize=\textwidth
\epsffile{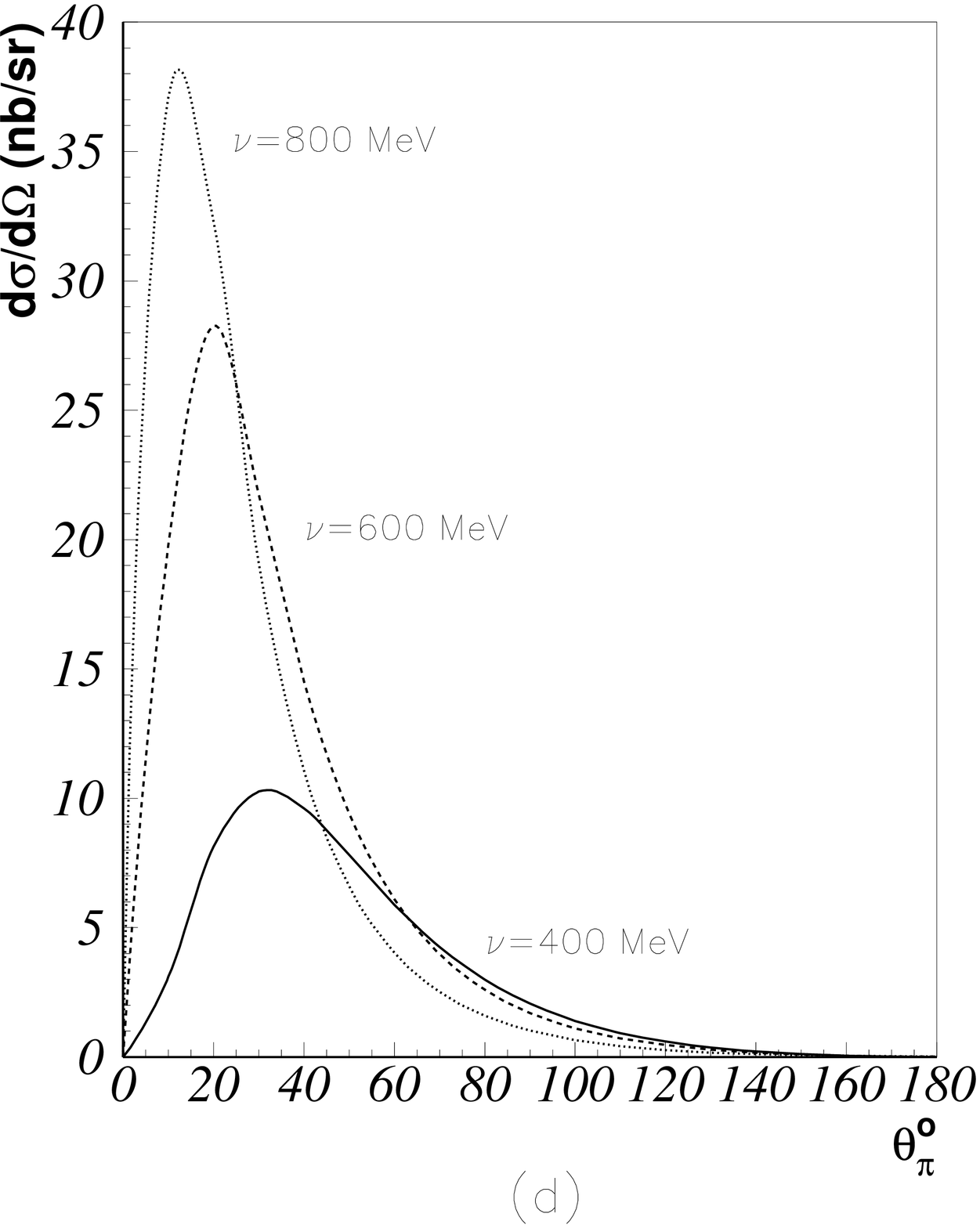}
\end{minipage}
\bigskip
\caption{The cross sections of the SND $D(1,1^-,0)$ production in the
reaction $\g d\to \pi^+D$; (a) --the total cross sections; (b,c,d) --
the differential cross sections for $M=1900$, 1942,, and 2000 MeV,
respectively.
\label{crsm}}
\end{figure}

\begin{figure}[ht]
\begin{minipage}{0.45\textwidth}   
\epsfxsize=\textwidth
\epsffile{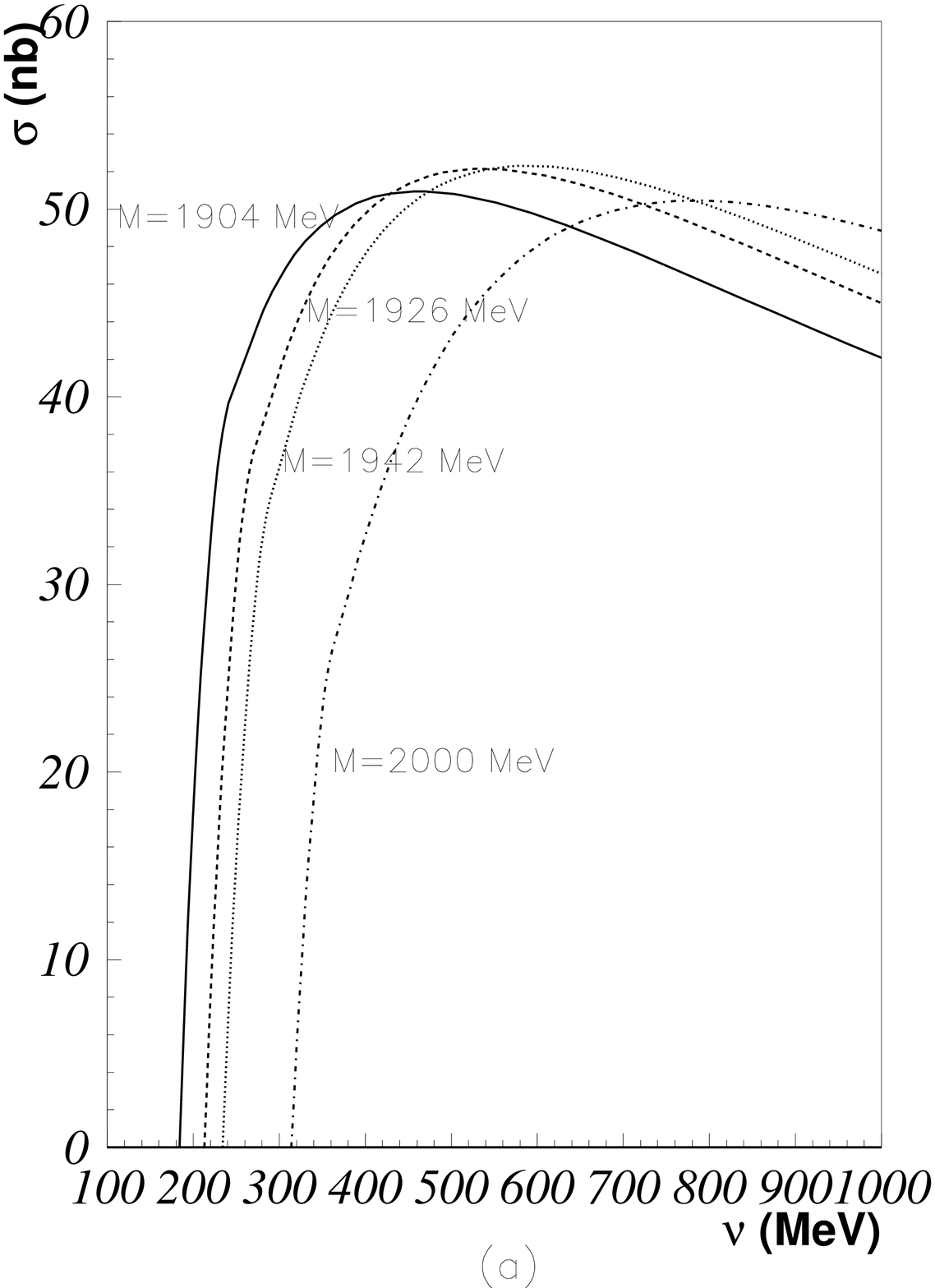}
\end{minipage}
\begin{minipage}{0.45\textwidth}
\epsfxsize=\textwidth
\epsffile{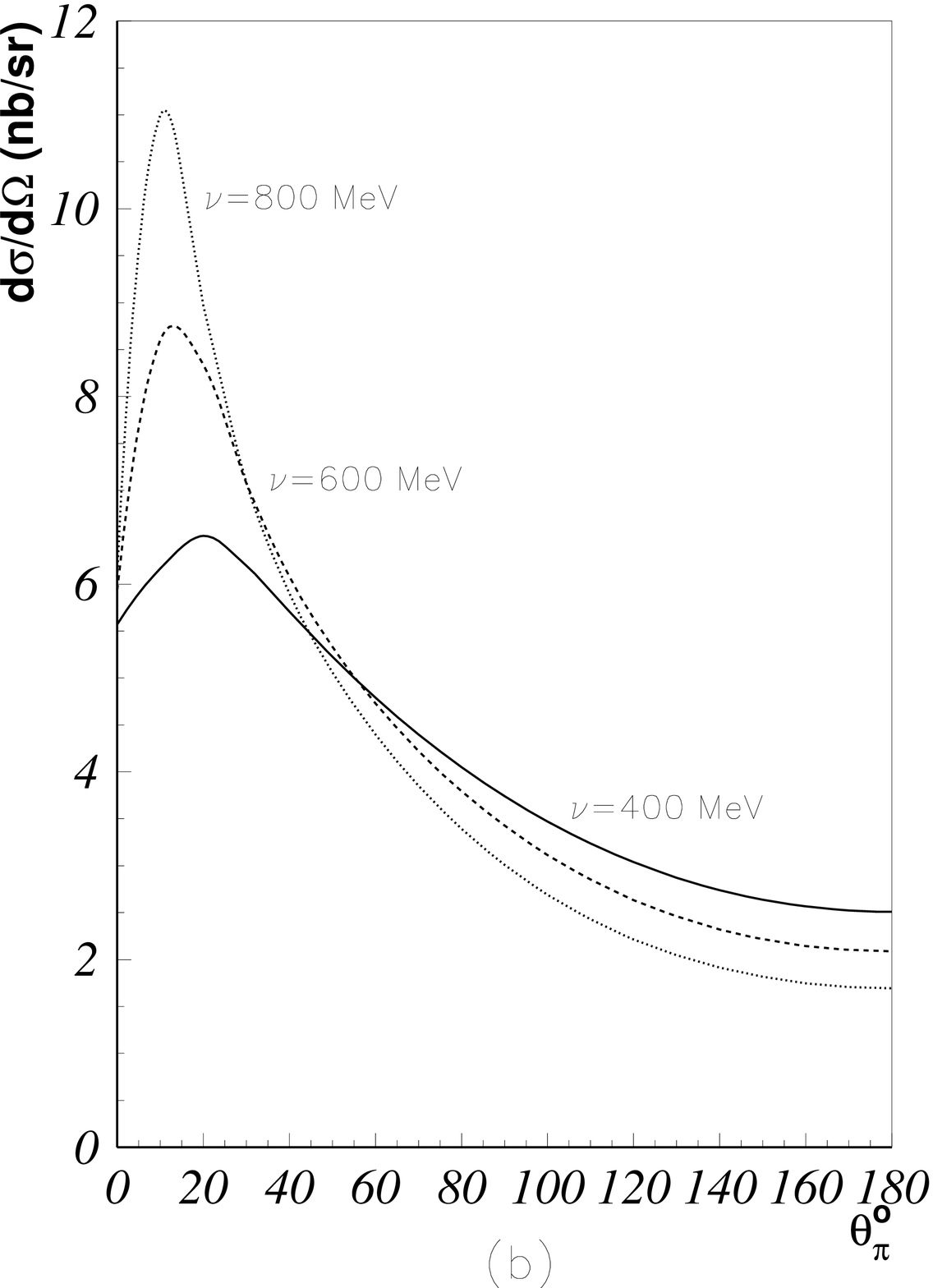}
\end{minipage}
\begin{minipage}{0.45\textwidth}
\epsfxsize=\textwidth
\epsffile{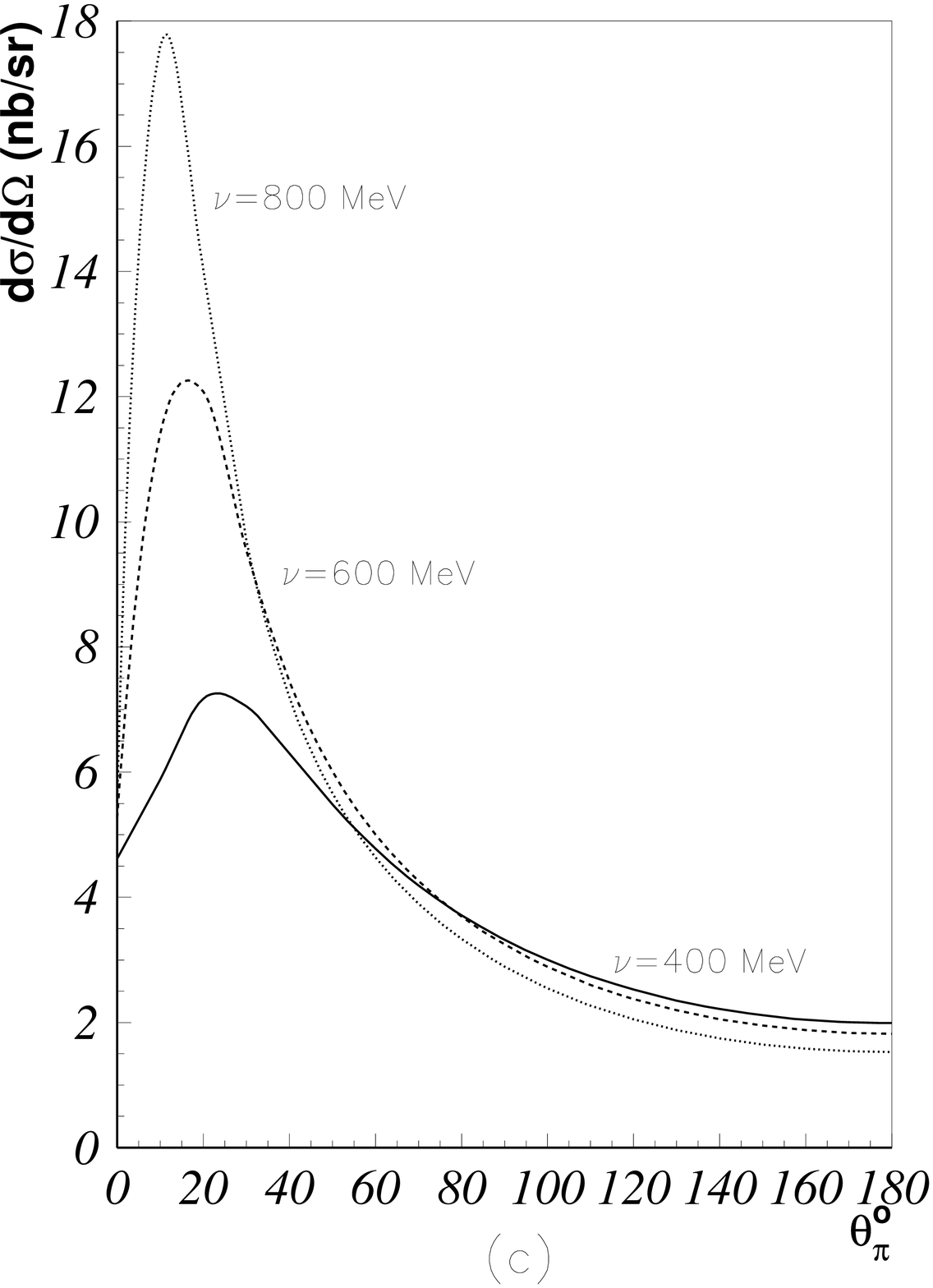}
\end{minipage}
\begin{minipage}{0.45\textwidth}
\epsfxsize=\textwidth
\epsffile{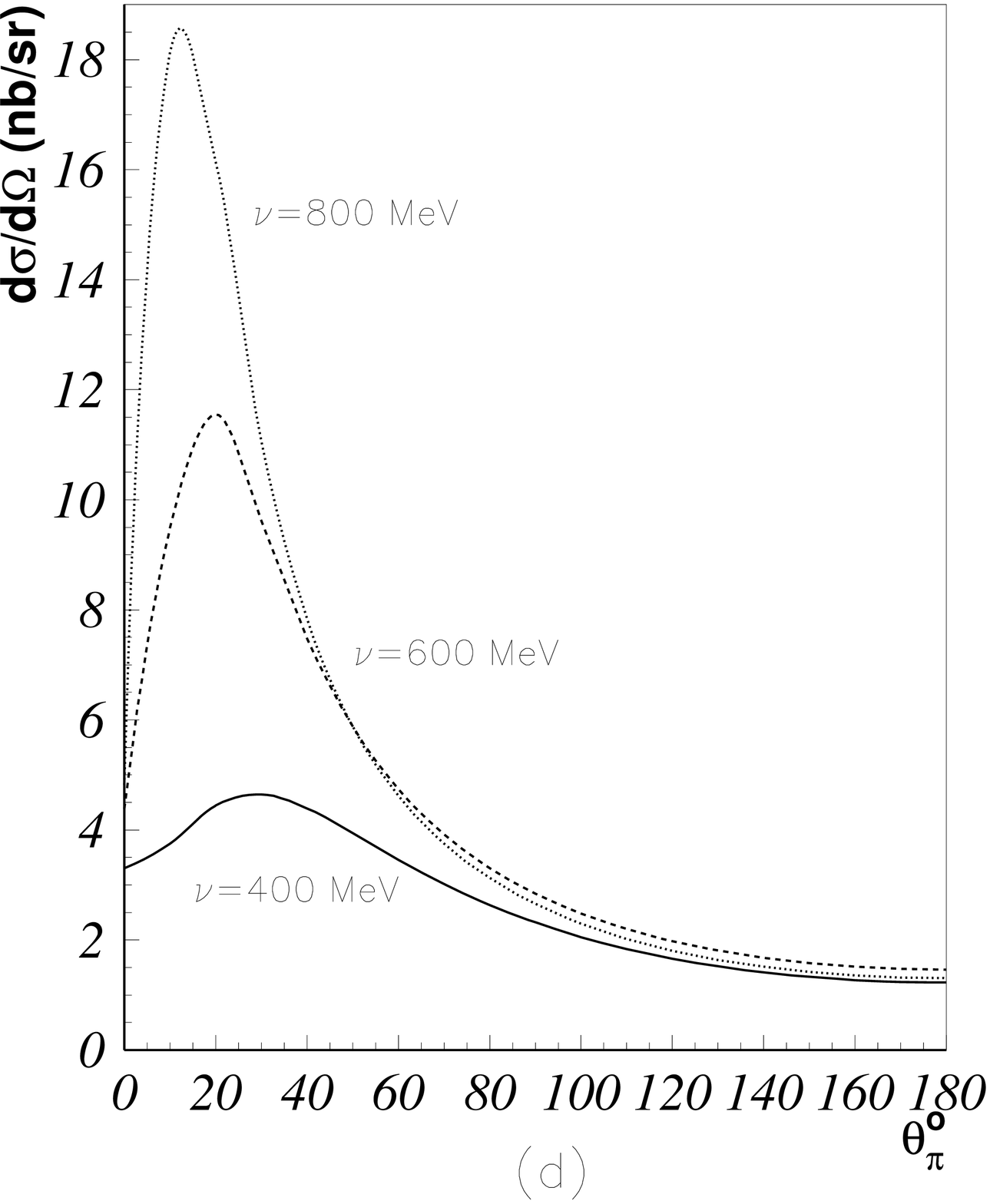}
\end{minipage}
\bigskip
\caption{The cross sections of the SND $D(1,1^+,1)$ production in the
reaction $\g d\to \pi^+D$; (a) --the total cross sections; (b,c,d) --
the differential cross sections for $M=1900$, 1942,, and 2000 MeV,
respectively.
\label{crsp}}
\end{figure}

\begin{figure}[ht]
\begin{minipage}{0.45\textwidth}
\epsfxsize=\textwidth
\epsffile{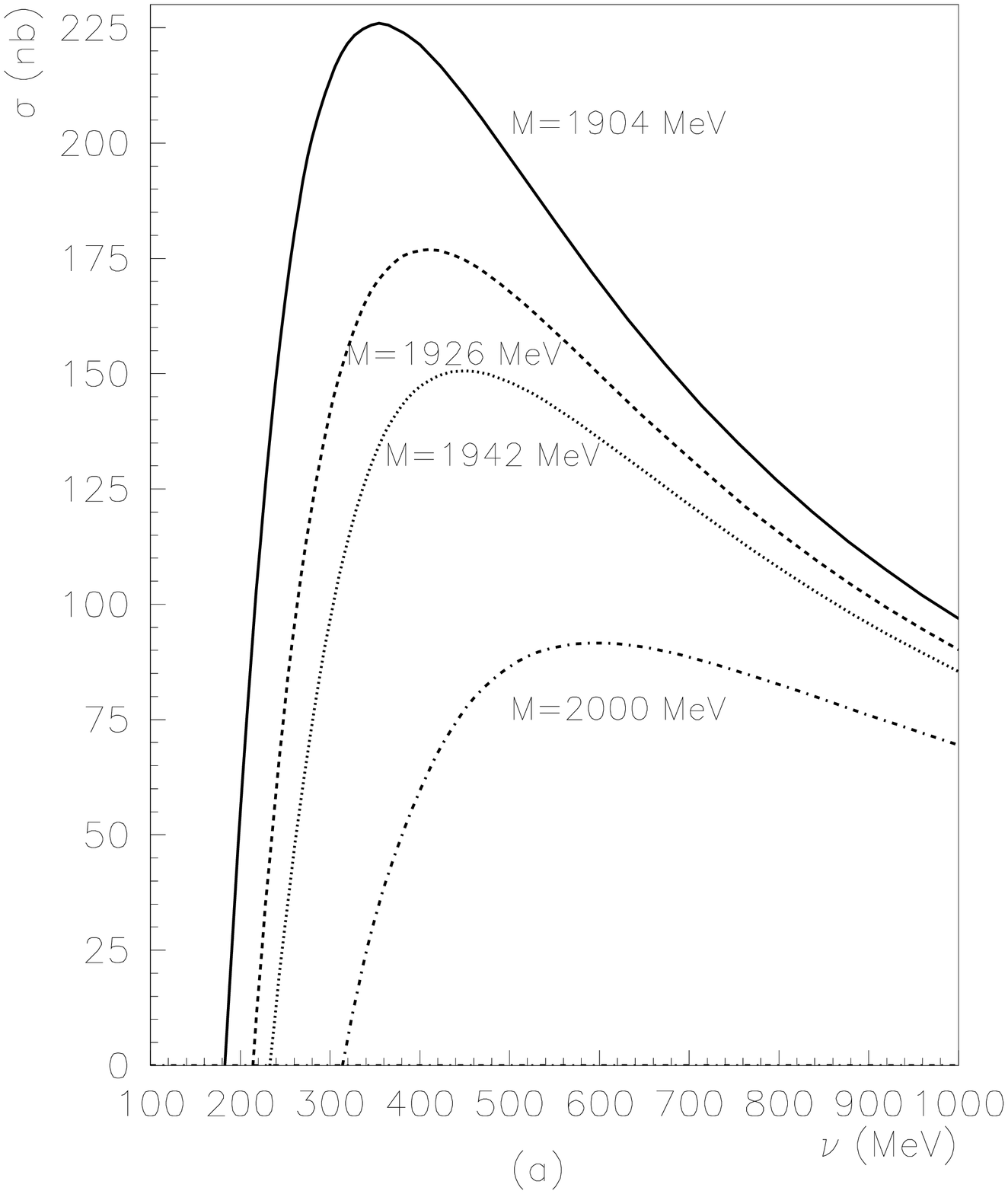}
\end{minipage}
\begin{minipage}{0.45\textwidth}
\epsfxsize=\textwidth
\epsffile{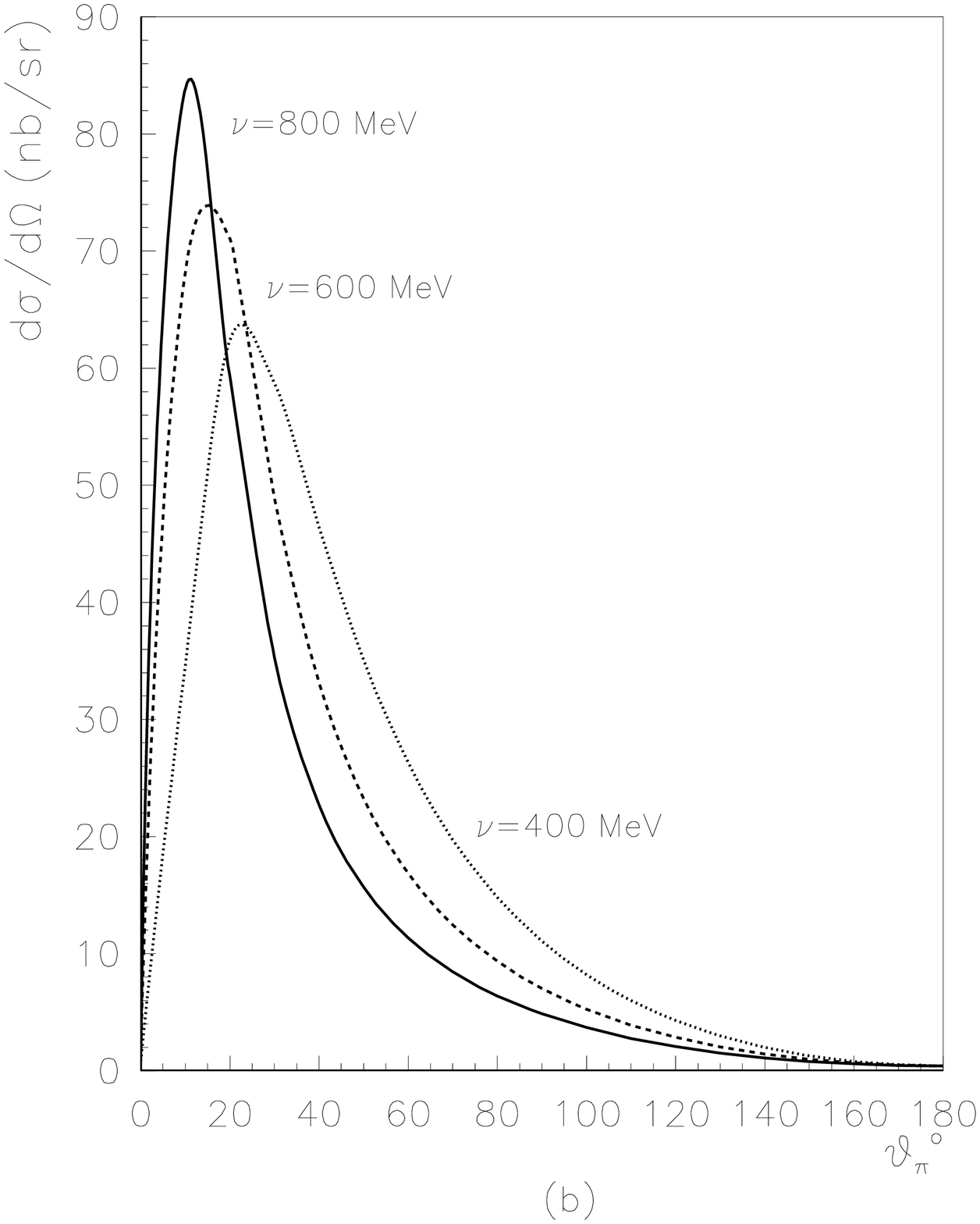}
\end{minipage}
\begin{minipage}{0.45\textwidth}
\epsfxsize=\textwidth
\epsffile{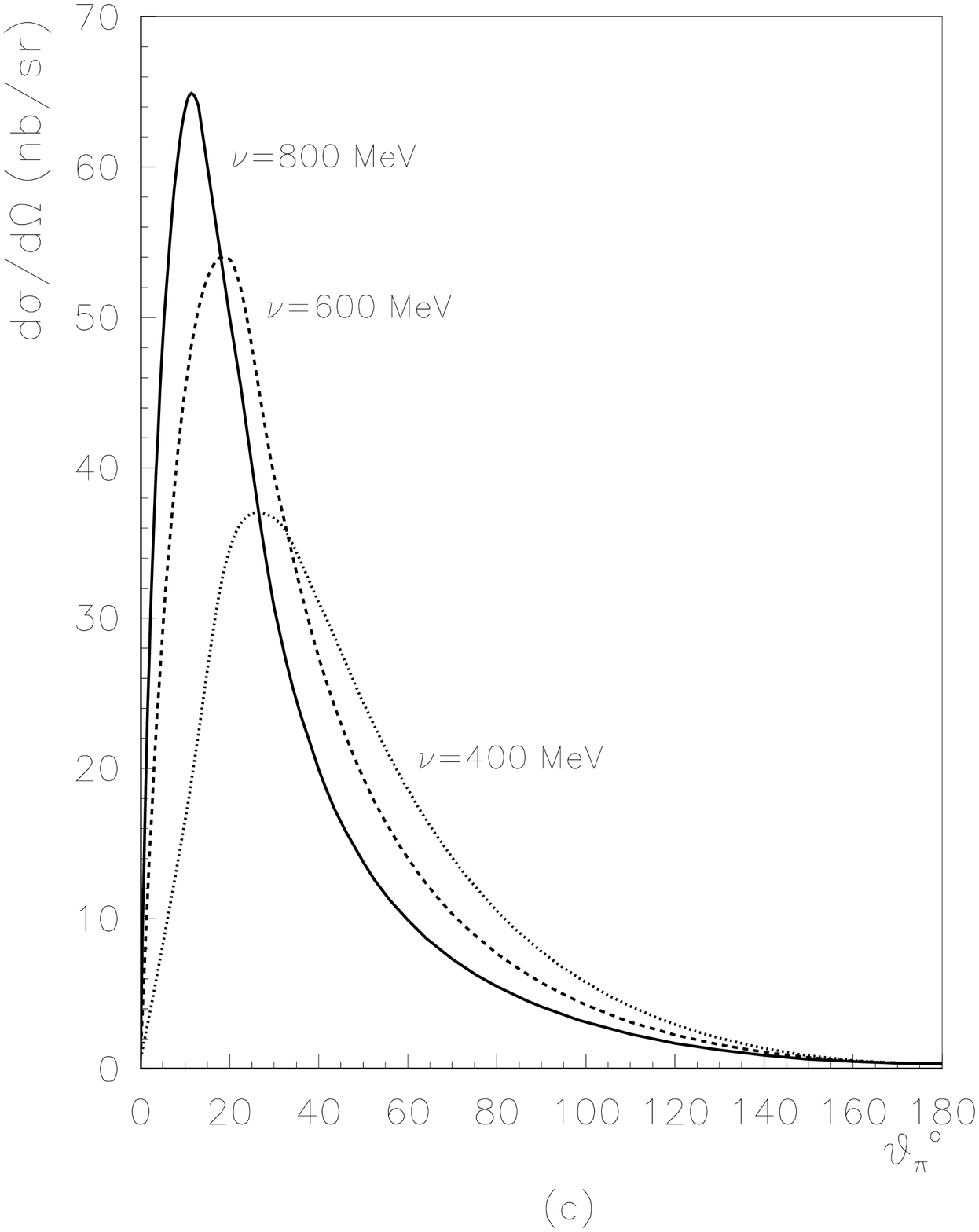}
\end{minipage}
\begin{minipage}{0.45\textwidth}
\epsfxsize=\textwidth
\epsffile{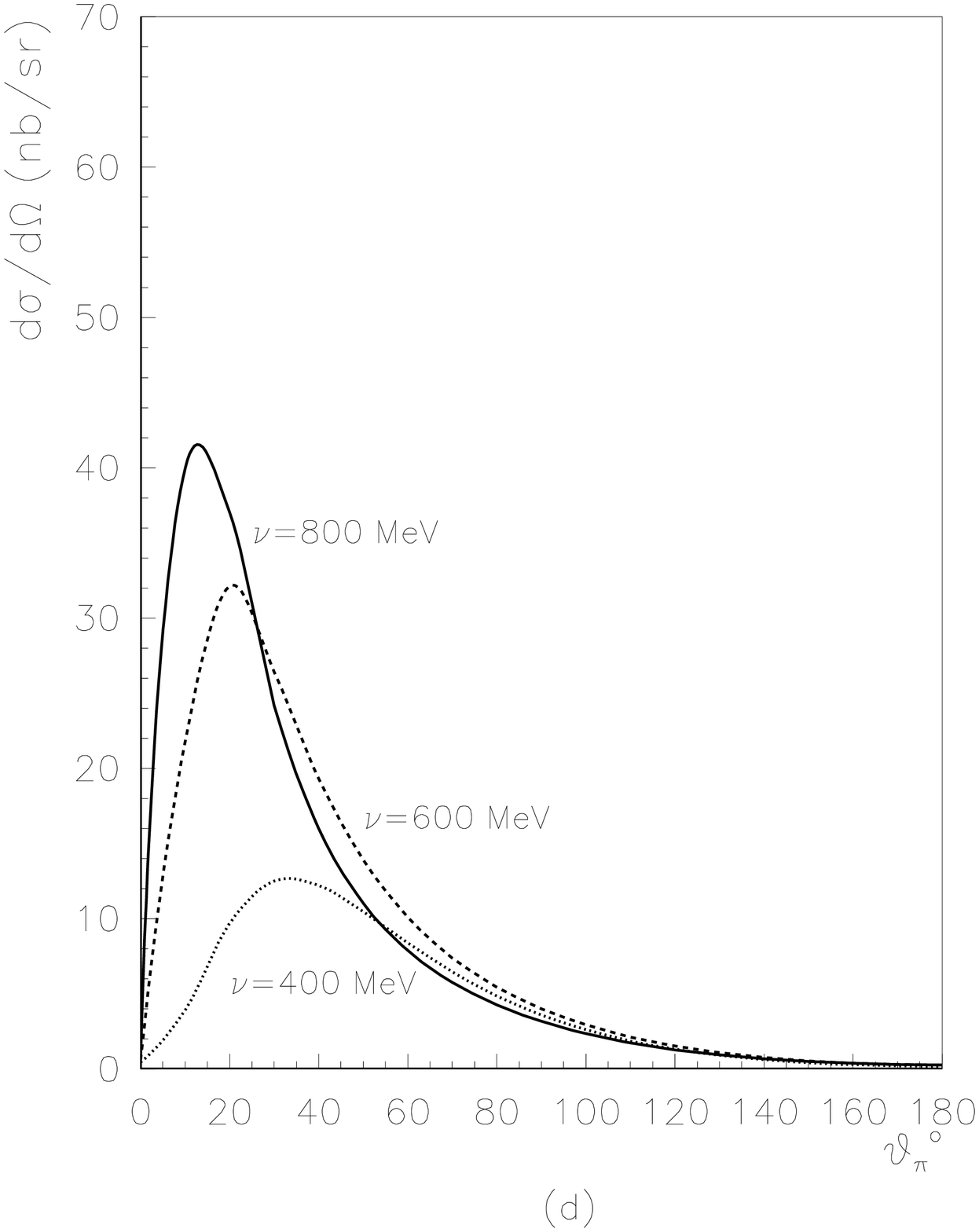}
\end{minipage}
\bigskip
\caption{The cross sections of the SND $D(1,1^-,0)$ production in the
reaction $\g d\to \pi^-D$; (a) --the total cross sections; (b,c,d) --
the differential cross sections for $M=1904$, 1942,, and 2000 MeV,
respectively.
\label{crsmm}}
\end{figure}

\begin{figure}[ht]
\begin{minipage}{0.45\textwidth}
\epsfxsize=\textwidth
\epsffile{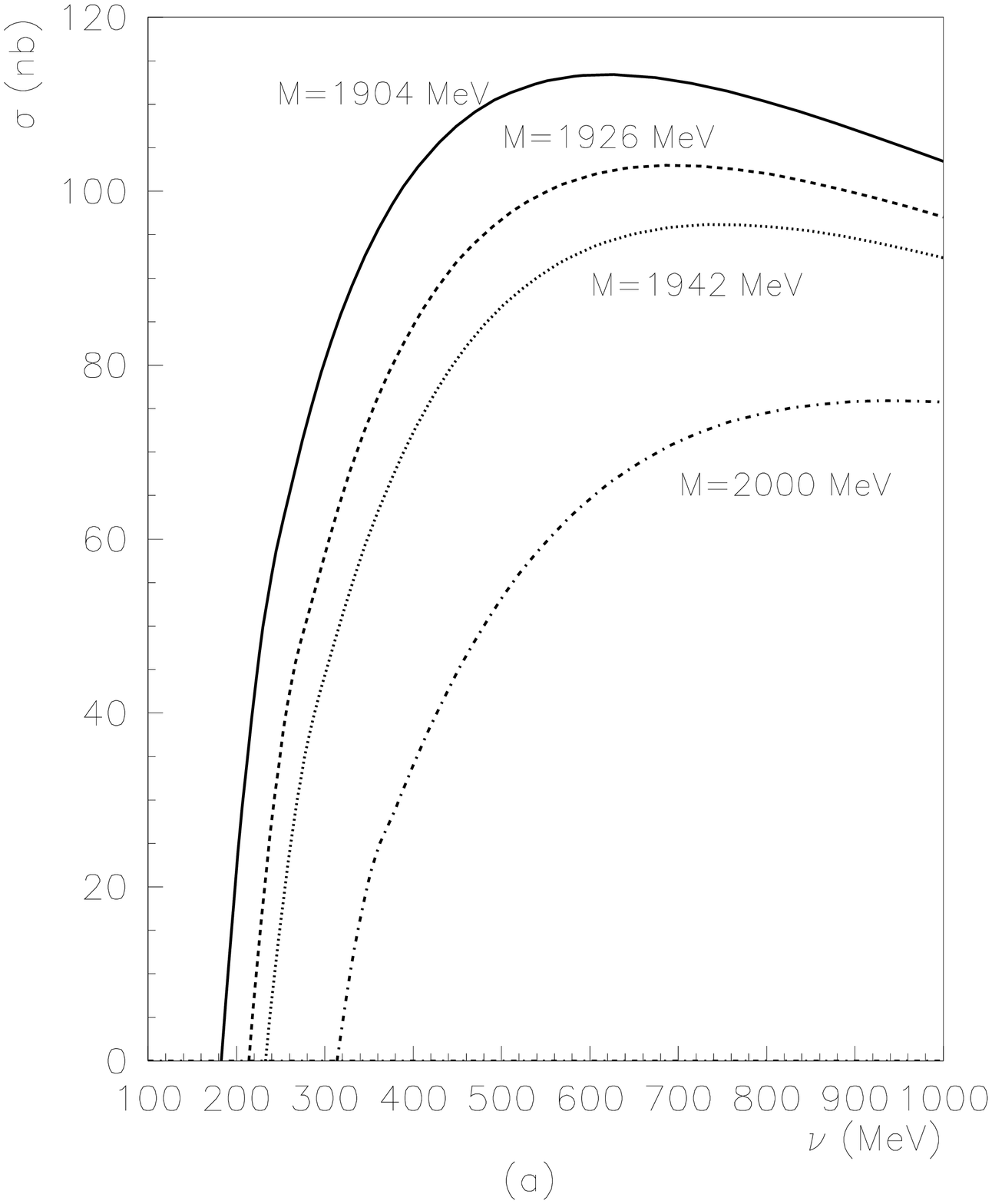}
\end{minipage}
\begin{minipage}{0.45\textwidth}
\epsfxsize=\textwidth
\epsffile{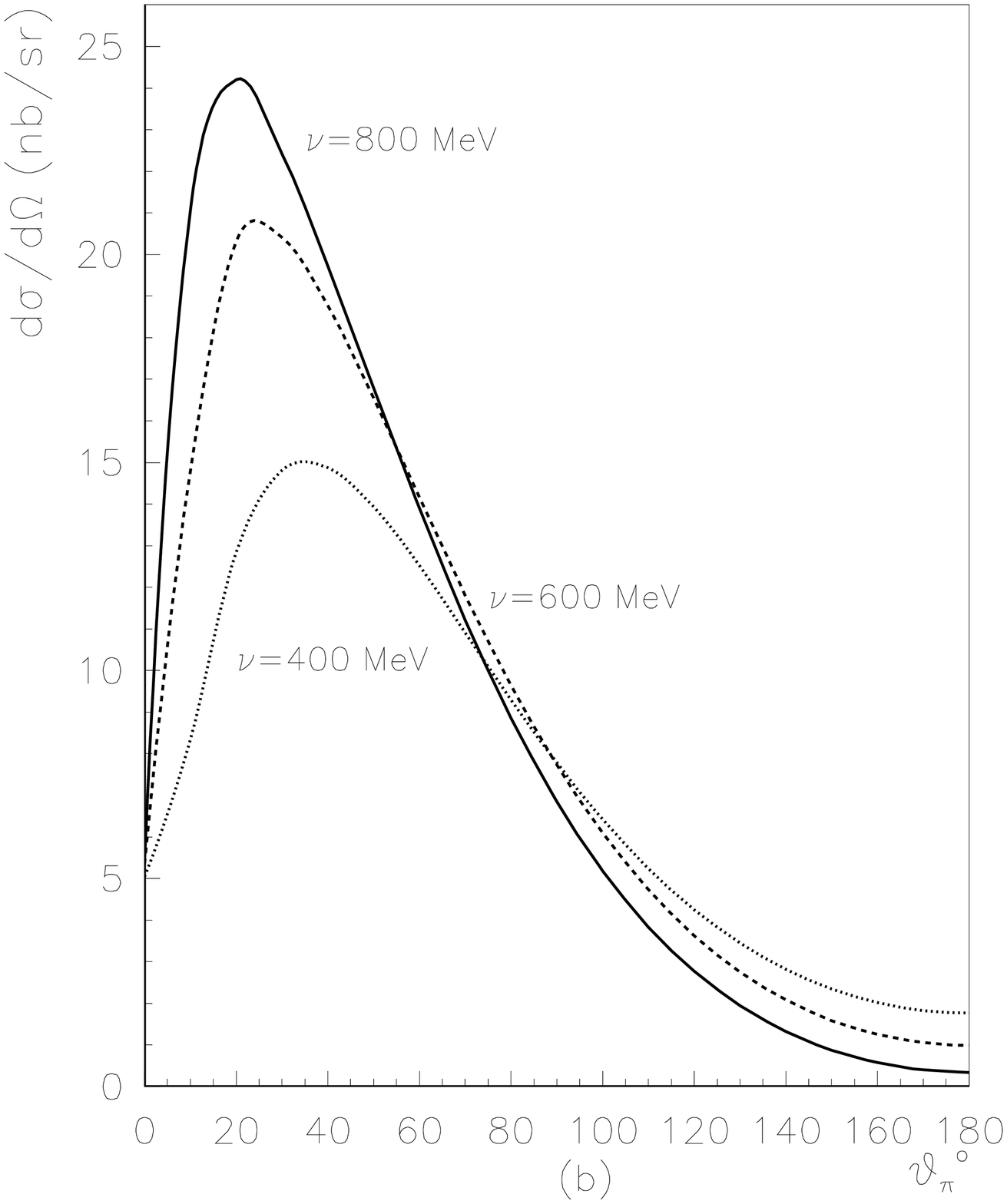}
\end{minipage}
\begin{minipage}{0.45\textwidth}
\epsfxsize=\textwidth
\epsffile{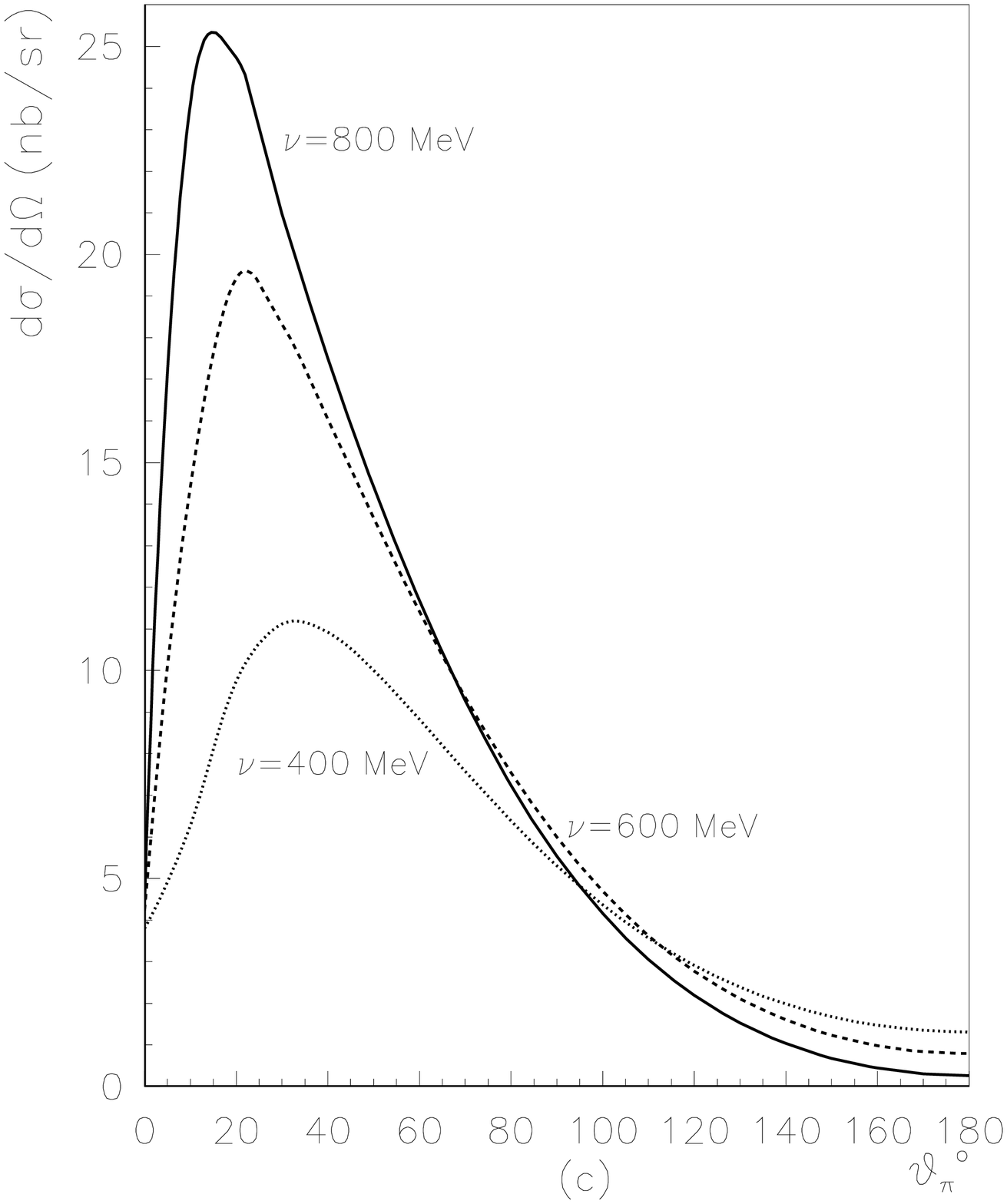}
\end{minipage}
\begin{minipage}{0.45\textwidth}
\epsfxsize=\textwidth
\epsffile{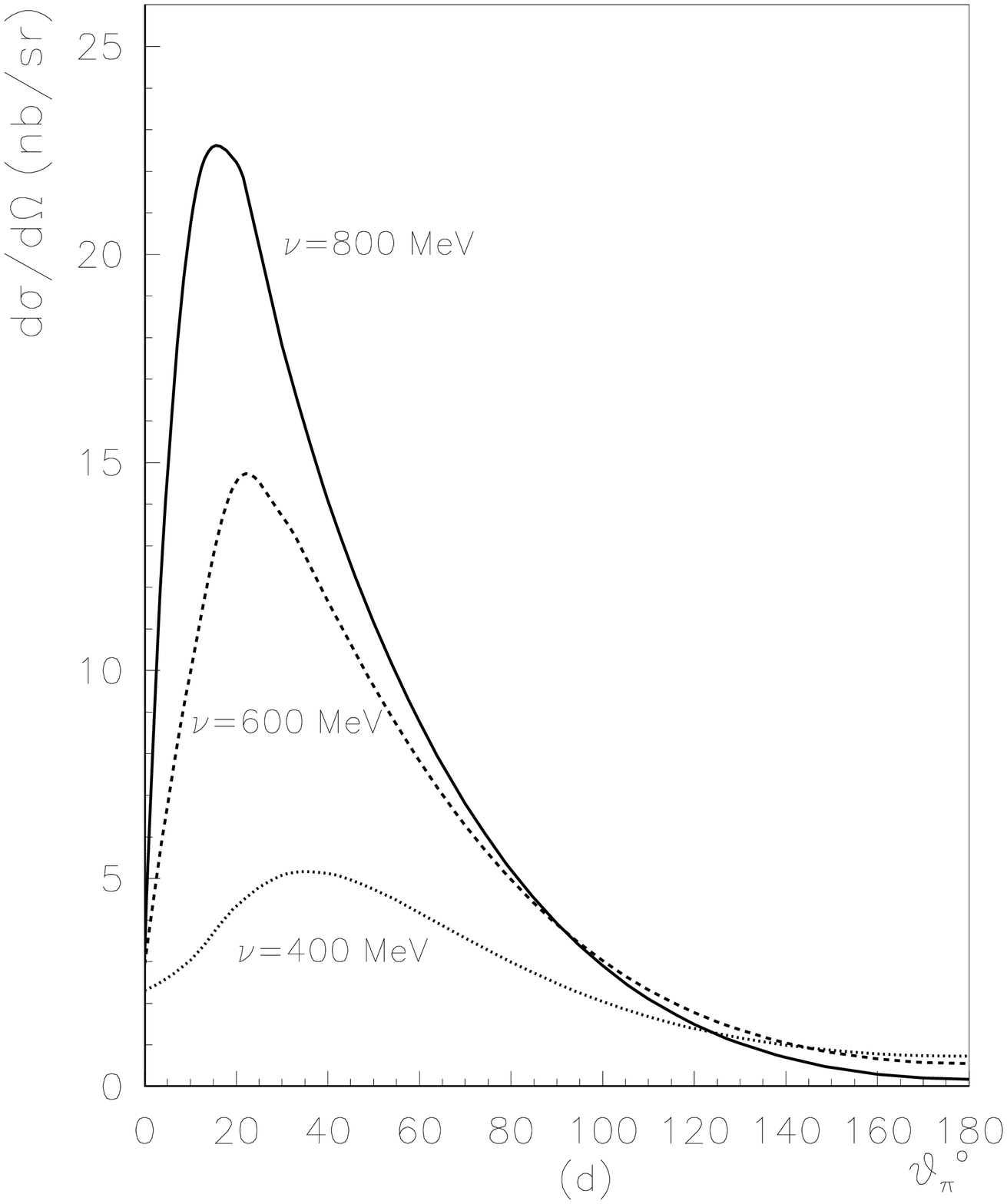}
\end{minipage}
\bigskip
\caption{The cross sections of the SND $D(1,1^+,1)$ production in the
reaction $\g d\to \pi^-D$; (a) --the total cross sections; (b,c,d) --
the differential cross sections for $M=1904$, 1942,, and 2000 MeV,
respectively.
\label{crspm}}
\end{figure}

\begin{figure}[ht]
\begin{minipage}{0.45\textwidth}
\epsfxsize=\textwidth
\epsffile{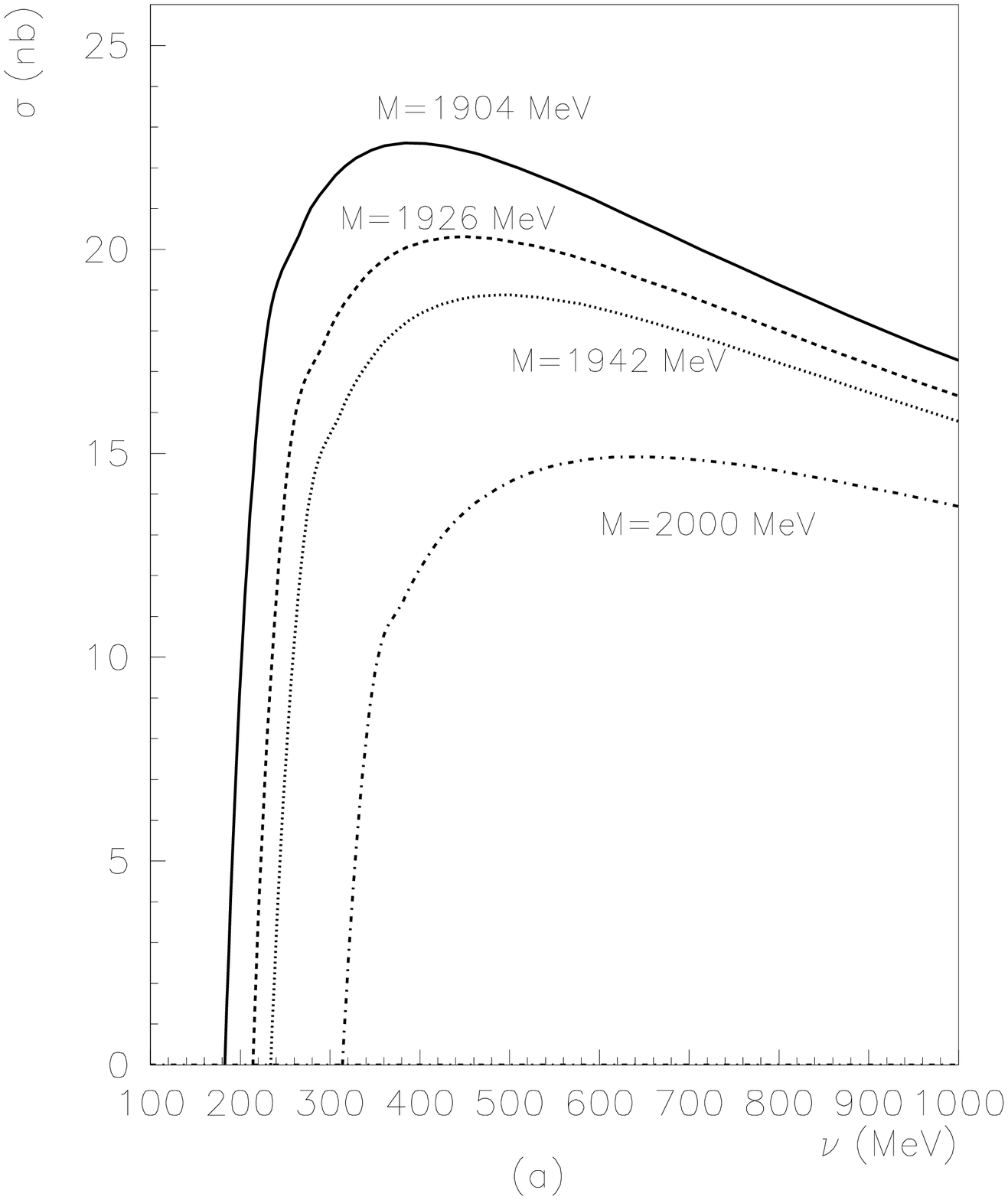}
\end{minipage}
\begin{minipage}{0.45\textwidth}
\epsfxsize=\textwidth
\epsffile{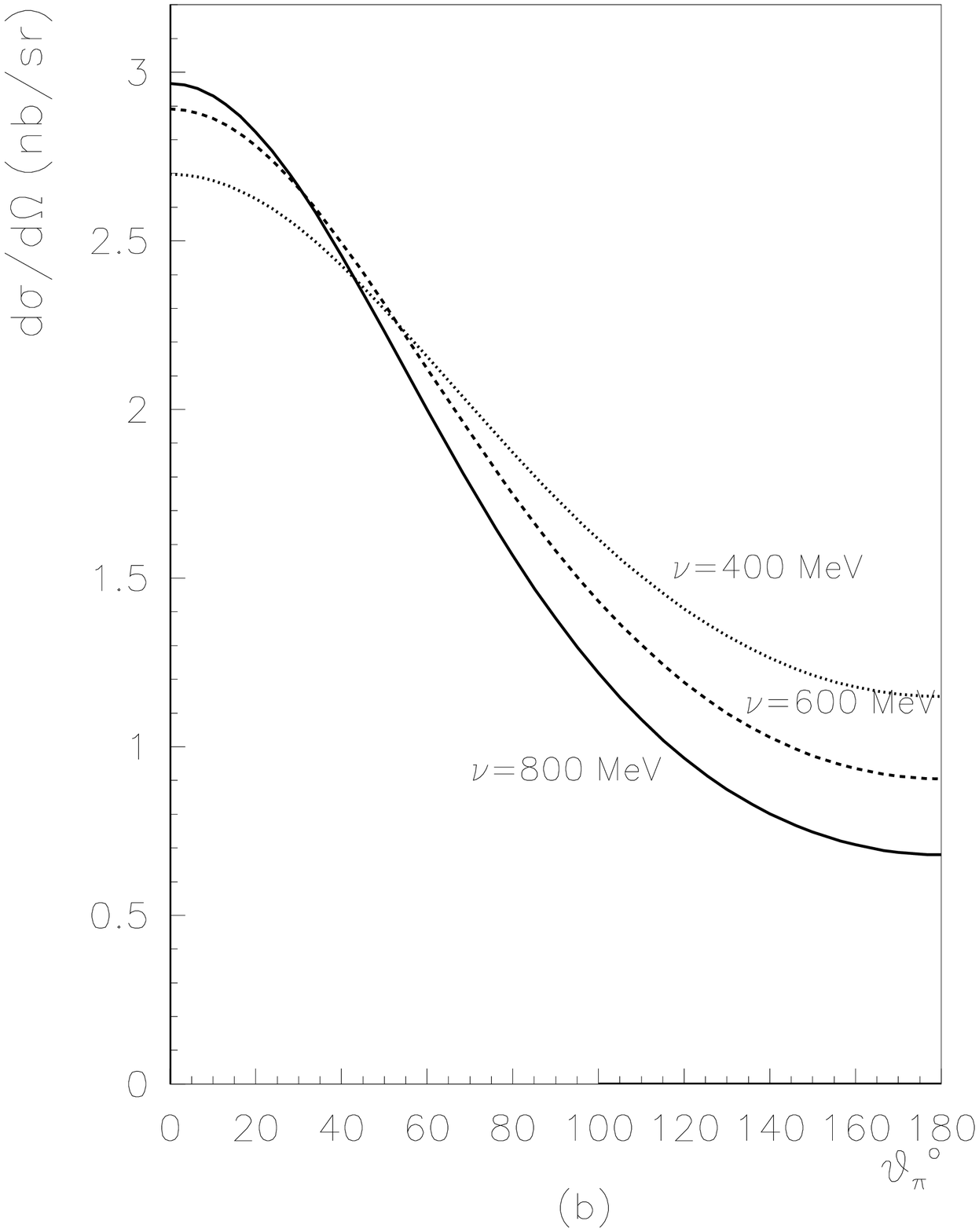}
\end{minipage}
\begin{minipage}{0.45\textwidth}
\epsfxsize=\textwidth
\epsffile{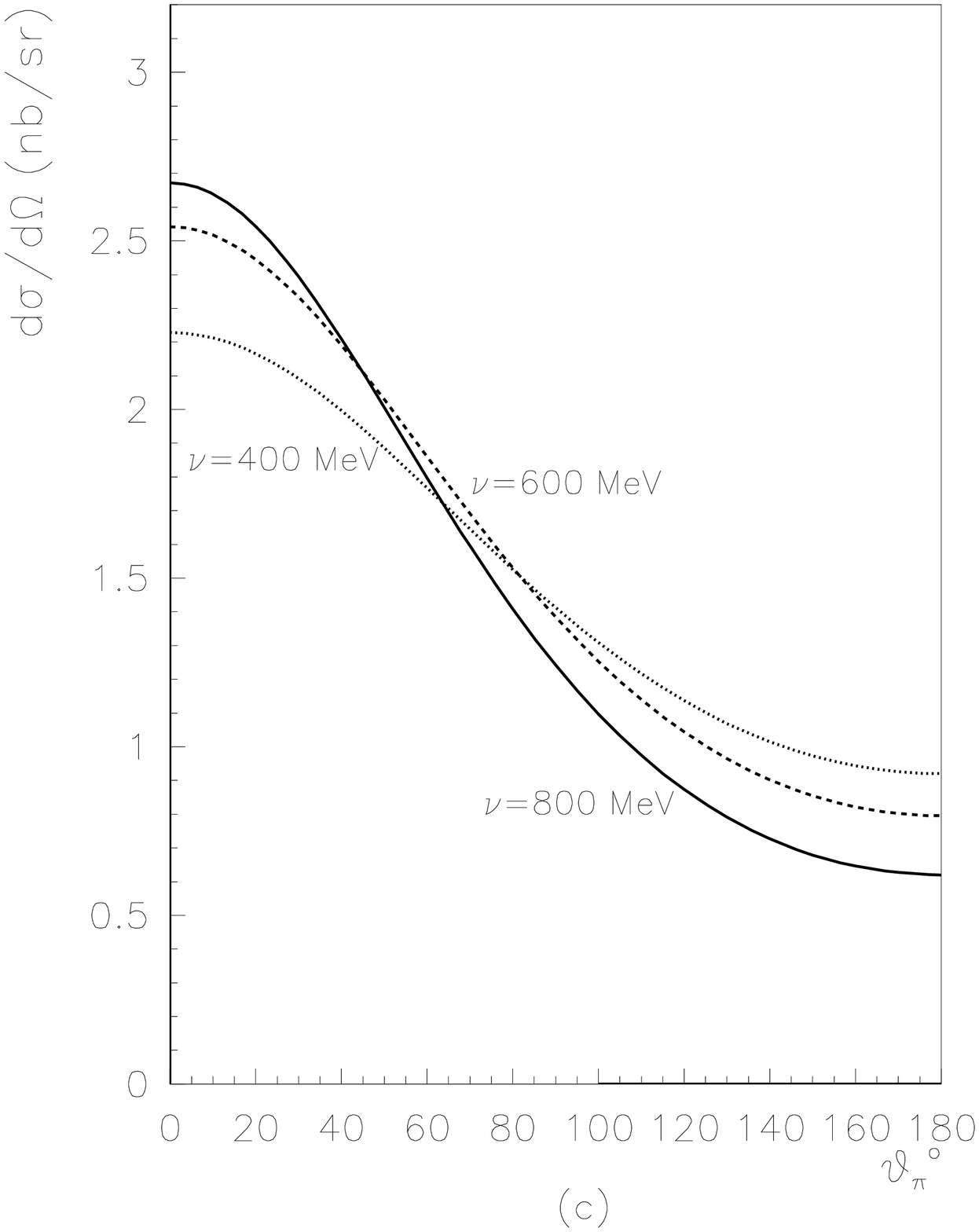}
\end{minipage}
\begin{minipage}{0.45\textwidth}
\epsfxsize=\textwidth
\epsffile{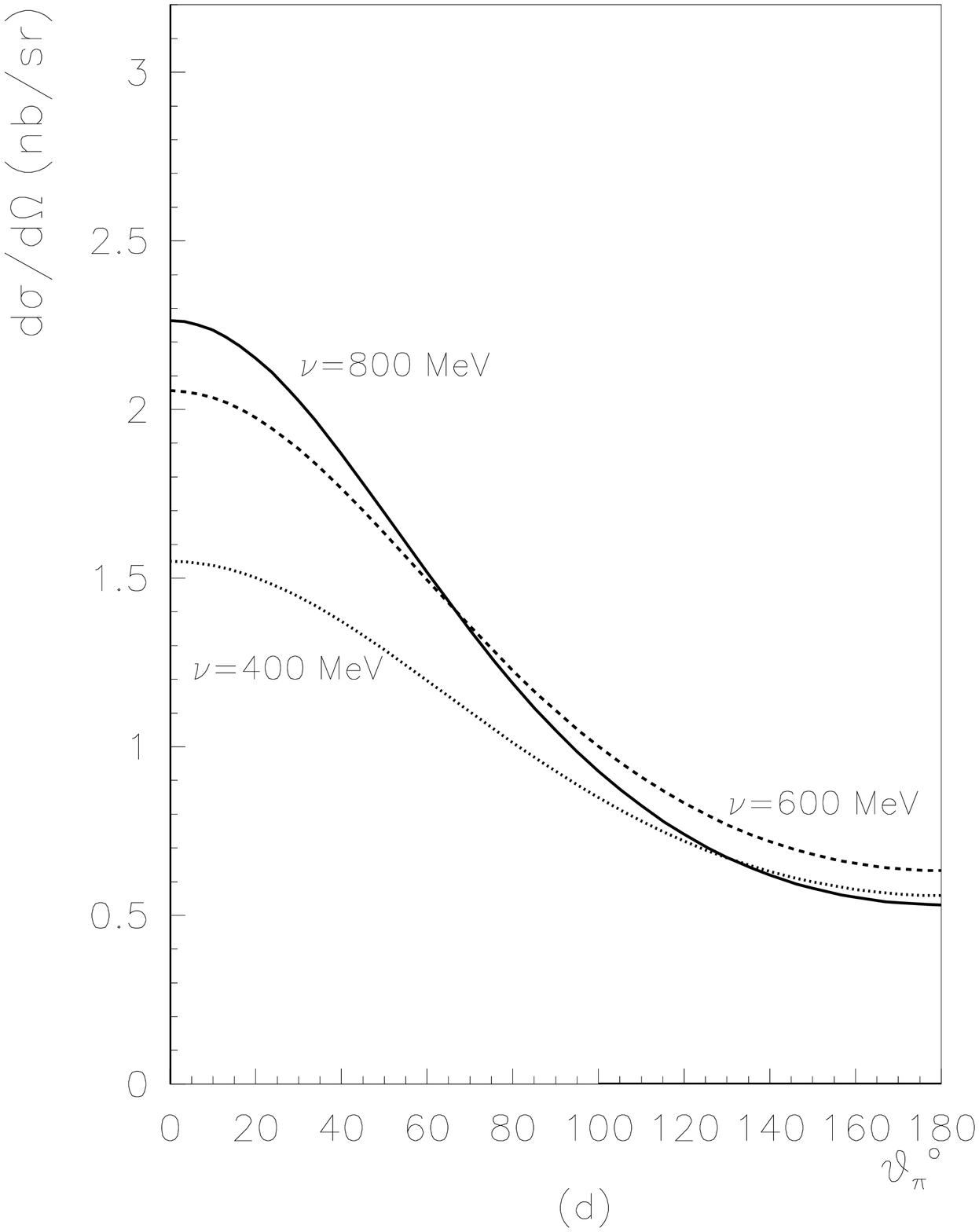}
\end{minipage}
\bigskip
\caption{The cross sections of the SND $D(1,1^+,1)$ production in the
reaction $\g d\to \pi^0D$; (a) --the total cross sections; (b,c,d) --
the differential cross sections for $M=1904$, 1942,, and 2000 MeV,
respectively.
\label{crsp0}}
\end{figure}

\begin{figure}[ht]
\begin{minipage}{0.45\textwidth}
\epsfxsize=\textwidth
\epsffile{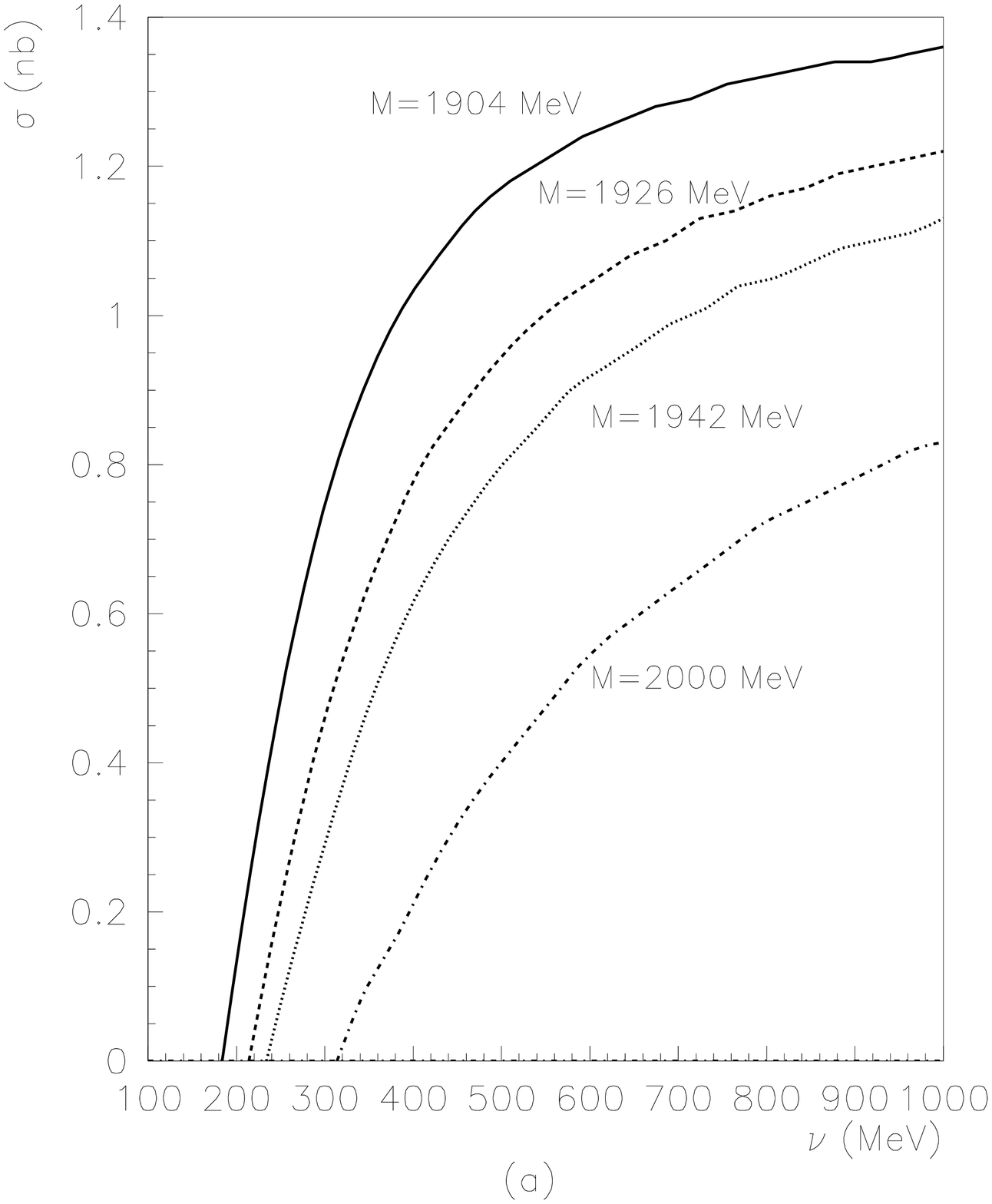}
\end{minipage}
\begin{minipage}{0.45\textwidth}
\epsfxsize=\textwidth
\epsffile{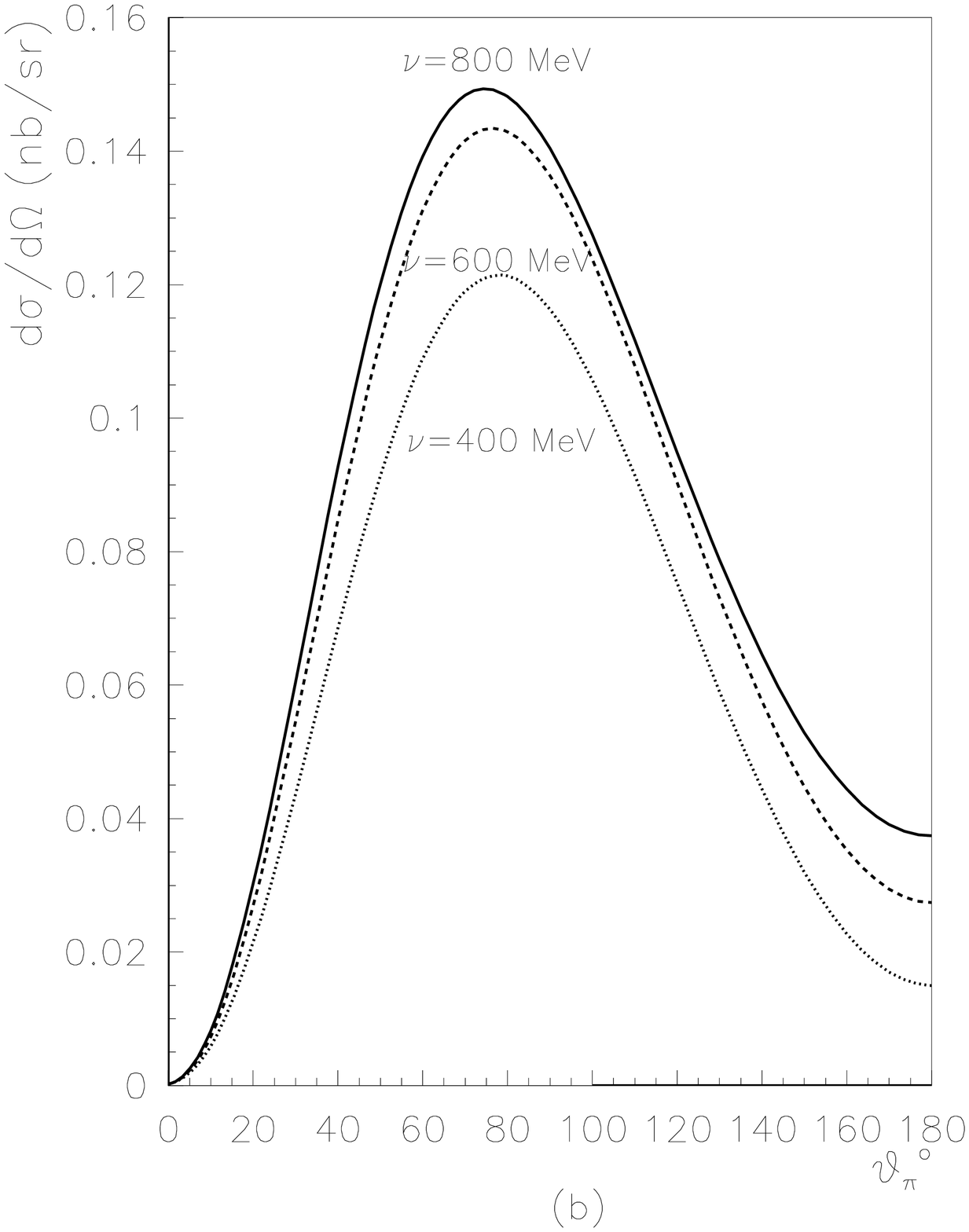}
\end{minipage}
\begin{minipage}{0.45\textwidth}
\epsfxsize=\textwidth
\epsffile{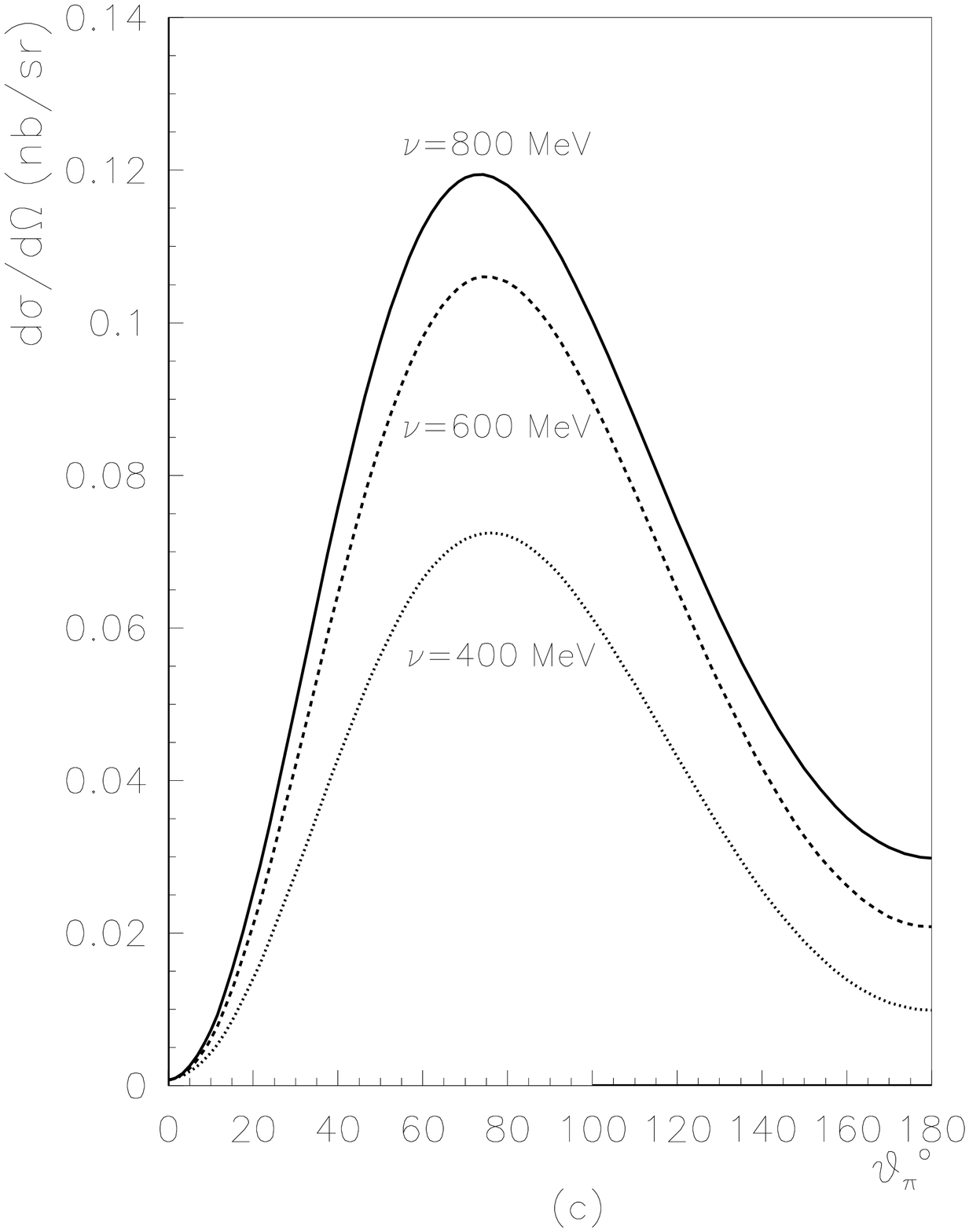}
\end{minipage}
\begin{minipage}{0.45\textwidth}
\epsfxsize=\textwidth
\epsffile{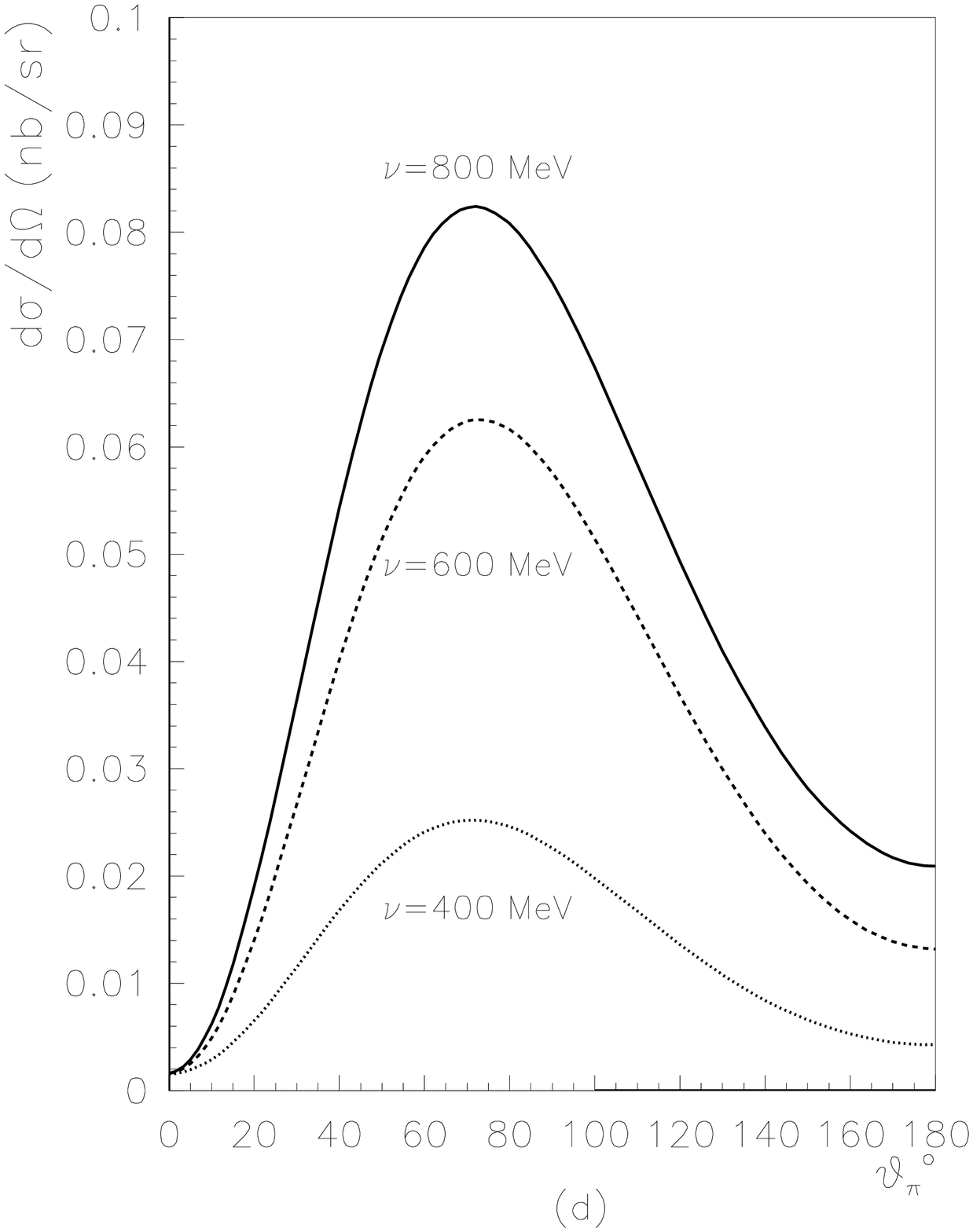}
\end{minipage}
\bigskip
\caption{The cross sections of the SND $D(1,1^-,0)$ production in the
reaction $\g d\to \pi^0D$; (a) --the total cross sections; (b,c,d) --
the differential cross sections for $M=1904$, 1942,, and 2000 MeV,
respectively.
\label{crsm0}}
\end{figure}

Results of the calculations of the total and differential cross sections
of the $D(1,1^{-},0)$ and $D(1,1^+,1)$ production in the processes of
the charged and neutral pion photoproduction from the deuteron
are presented in Figs. \ref{crsm} -- \ref{crsm0}.

The total cross sections of the
$D(1,1^-,0)$ production in the reaction with the $\pp$ meson formation
(Fig. \ref{crsm}) change from 50 nb up to 160 nb in the photon energy
interval under consideration. The main contribution to the differential
cross sections is given by the pions emitted to the angular region
$0^{\circ}\div 100^{\circ}$ with maximum at $10^{\circ}-30^{\circ}$.

The total cross sections of the $D(1,1^+,1)$ photoproduction in this
reaction are equal to $\sim 50$ nb (Fig. \ref{crsp}).

In the case of the $\pn$ meson photoproduction, the total cross sections
of the $D(1,1^-,0)$ and $D(1,1^+,1)$ production are bigger and reach for
$M=1904$ MeV to 225 nb and 110 nb, respectively (Fig. \ref{crsmm},
\ref{crspm}).

The values of the total cross sections of the $D(1,1^+,1)$ production
in the reaction $\g d\to\po D$ are significantly less and are equal to
15 -- 22 nb in the maximum (Fig. \ref{crsp0}).
However, the detection efficiency of the
$\po$ meson photoproduction by the Crystal Ball Spectrometer and TAPS
is higher than it for the charged pions and, therefore, the good yields
are expected in this case. The angular distribution of the $\po$ mesons
for this process is wider and occupies a region from $0^{\circ}$ to
$180^{\circ}$. The cross sections of $D(1,1^-,0)$
production in the reaction under consideration are less by one order
(Fig. \ref{crsm0}).

So, the comparison of the SND photoproduction cross sections in the
reactions with the formation of $\pp$, $\pn$, and $\po$ mesons
allows the quantum numbers of the SND to be determined.

\begin{figure}[ht]
\centerline{
\epsfxsize=16cm
\epsfysize=20cm
\epsffile{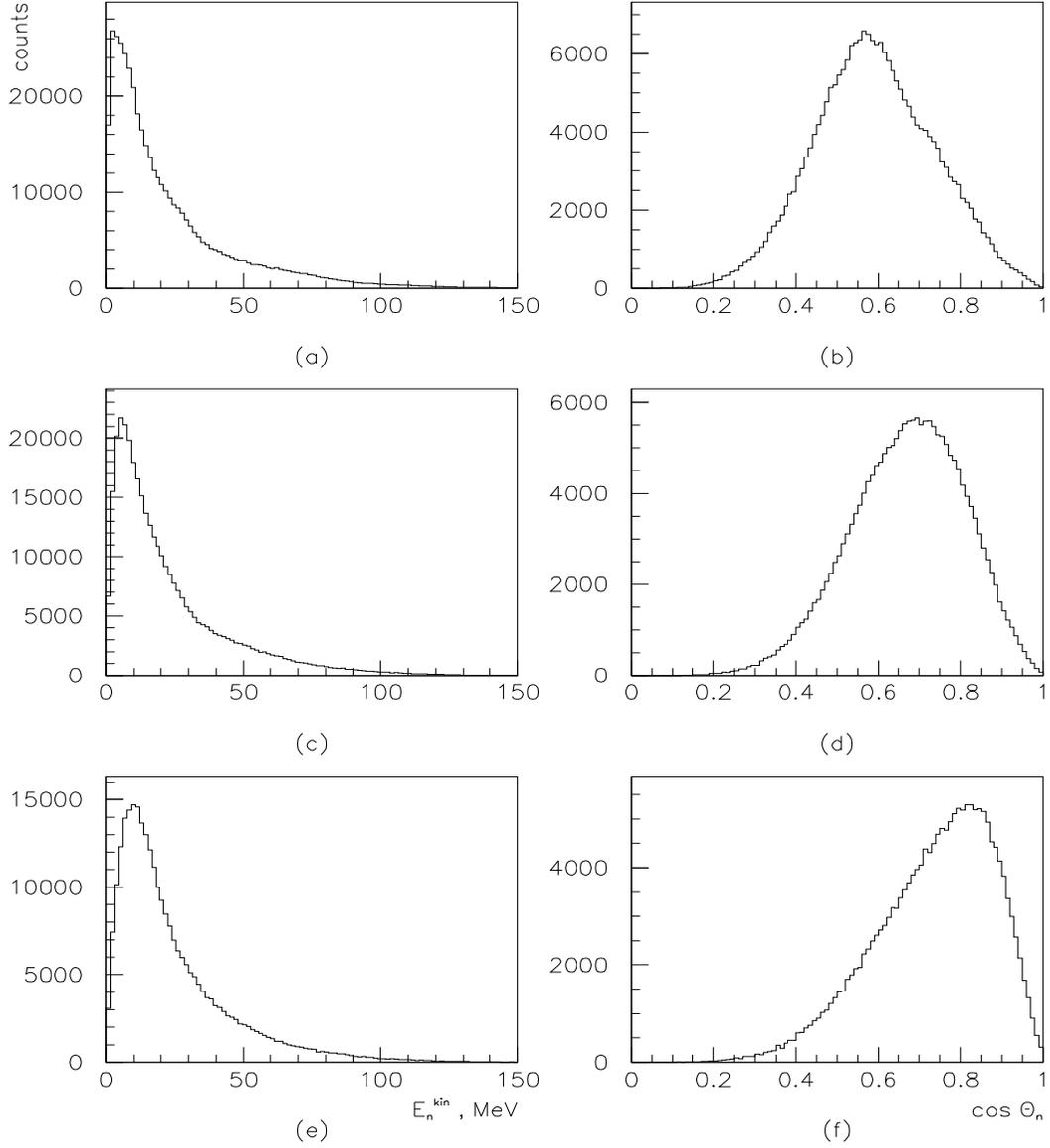}}
\caption{The energy (a,c,e) and angular (b,d,f) distributions of
the nucleons from the decays of the dibaryons with different masses:
(a,b) -- $M=1900$ MeV, (c,d) -- $M=1950$ MeV, (e,f) -- $M=2000$ MeV.}
\label{neut}
\end{figure}

\begin{figure}[ht]
\centerline{
\epsfxsize=16cm
\epsfysize=20cm
\epsffile{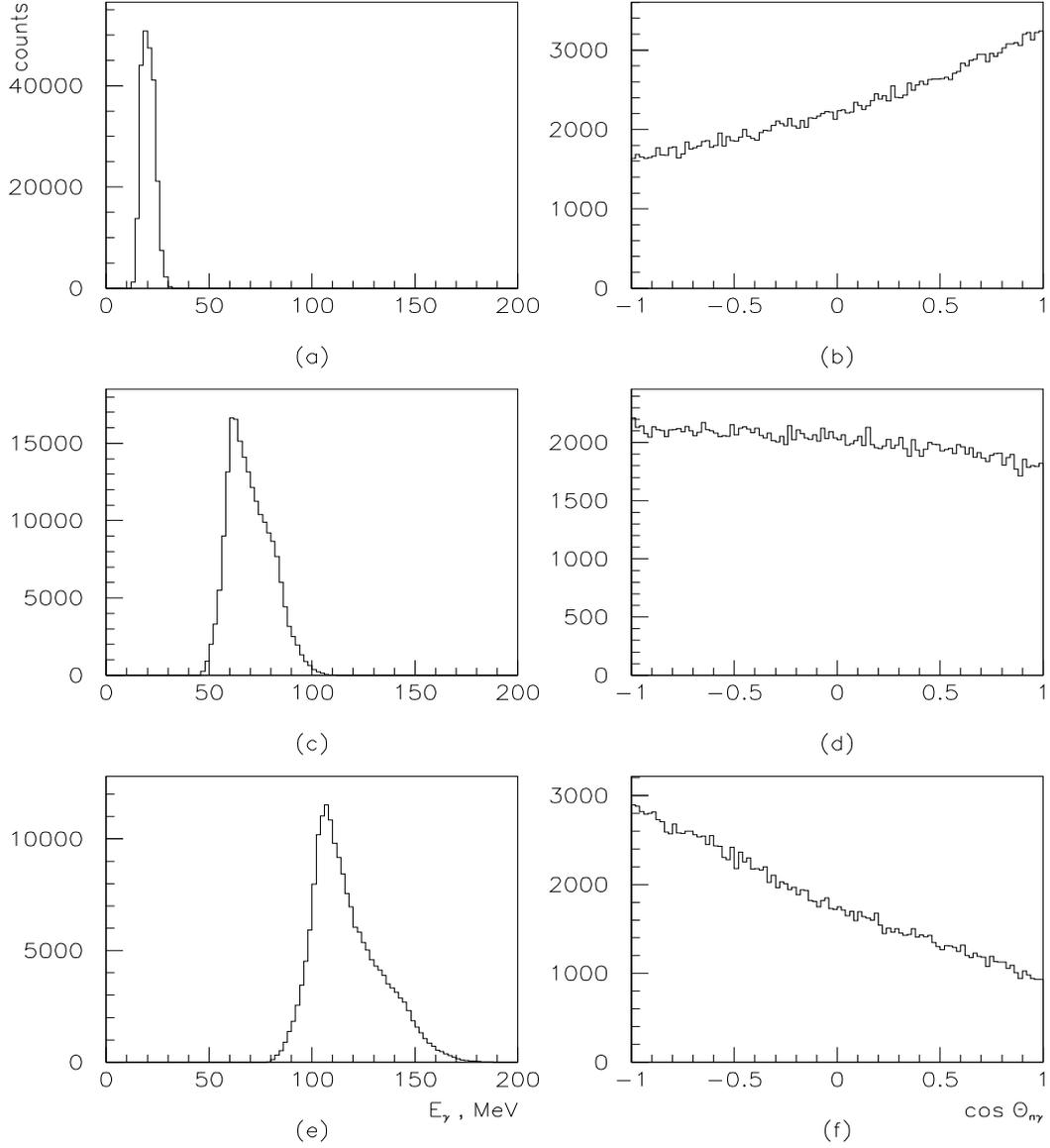}}
\caption{The energy (a,c,e) and angular (b,d,f) distributions of the
photons from the decays of the SND with the different masses:
(a,b) -- $M=1900$ MeV, (c,d) -- $M=1950$ MeV, (e,f) -- $M=2000$ MeV.}
\label{phot}
\end{figure}

Distributions of kinematical variables for the reaction under study
are presented in Figs. \ref{neut}, \ref{phot}
for the SND masses 1900, 1950 and 2000 $MeV$.

Fig. \ref{neut} demonstrates the expected distributions of the nucleons
over the energy and the emission angle. As seen from Fig. \ref{neut}a,c,e,
the nucleon energy is mainly smaller than 100 MeV. Therefore, we will
limit ourselves by a consideration of the nucleon with such an energy.
It will permit us to suppress the background essentially. The distribution
over $\cos\theta_n$ in Fig. \ref{neut}b,d,f
corresponds to an interval of the nucleon energy from 10 to 100 MeV and
shows that nucleons will be emitted mainly to the angles
$0^{\circ}-75^{\circ}$.

Fig. \ref{phot} demonstrates
distributions over the energy of the photon from the
dibaryon decay and over $\cos\theta_{\g n}$, where $\theta_{\g n}$ is
an angle between the nucleon and the final photon. These distributions were
obtained also for the nucleon energy in the interval 10 -- 100 MeV.

\section{Experimental setup}

\begin{figure}[ht]
\centerline{
\epsfxsize=16cm
\epsfysize=16cm
\epsffile{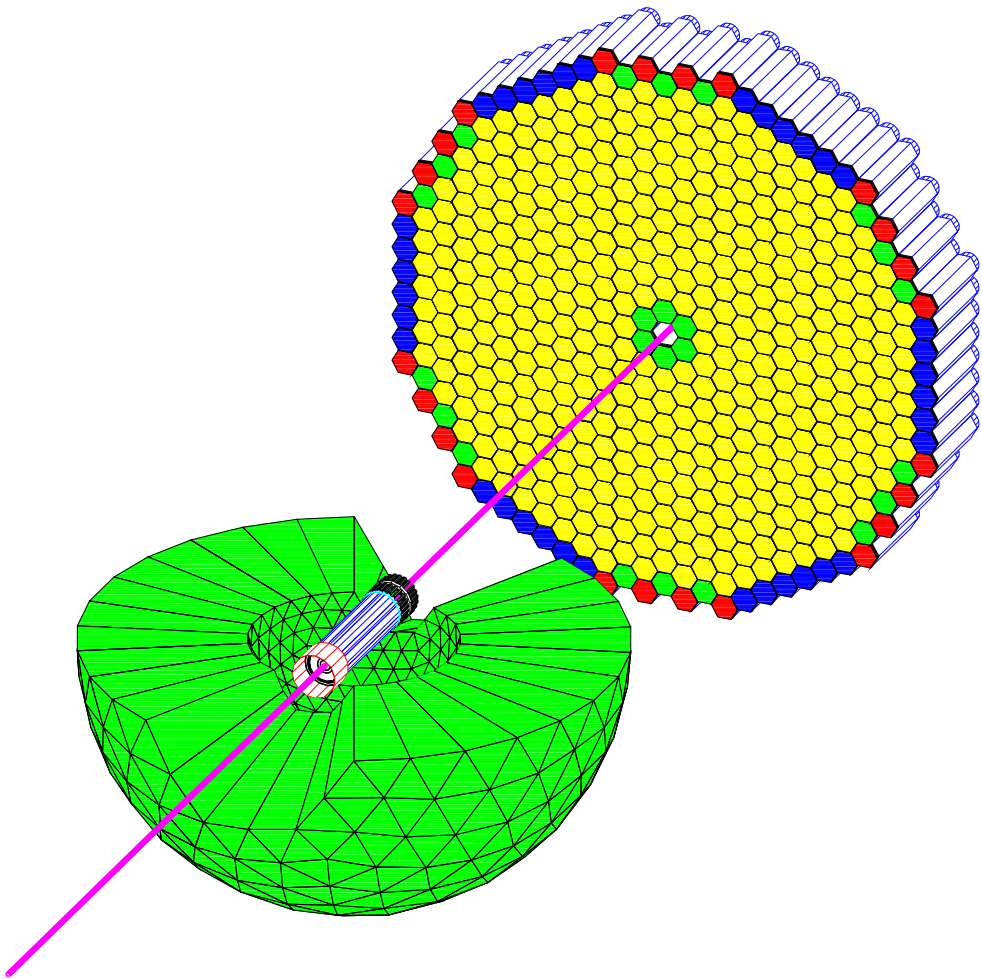}}      
\caption{Crystal Ball spectrometer and TAPS}
\label{cbt}
\end{figure}

We propose to use the Crystal Ball (CB) spectrometer and TAPS detector system
(CBTAPS) for the detection of all particles which are needed for the
identification of the narrow six-quark states.
The 5 cm long liquid hydrogen or deuterium target surrounded by inner
detectors is located inside the CB.
Experimental setup is shown schematically in Fig. \ref{cbt}. To show
the inner detectors, the upper hemisphere of the CB is omitted.

The CB is constructed of 672 NaJ(Tl) crystals with 15.7 radiation lenght
thick. The crystals are arranged in a spherical shell with an inner
radius of 25.3 cm and an outer radius of 66.0 cm. The CB has an entrance
and exit opening for the beam.

The inner detectors include two coaxial cylindrical multiwire proportional
chambers (MWPC), developed early for the DAPHNE detector \cite{mwpc}, and
a cylindrically arranged plastic scintillator particle-identification
detector (PID) \cite{pid} placed between the MWPC and the target.
The MWPC will be used as a charge particle tracker. The PID will be used,
from one side, as a $\Delta E$-detector to separate pions and protons and,
from other side, as a "Veto" for charged particles when photons
in the CB are detected.

The TAPS detector \cite{taps1,taps2} is implemented as a forward
wall at 1.8 m from the target. It comprises 510 BaF$_2$
hexagonally shaped crystals with 12 radiation length thick.
A hexagonal plastic scintillator detector (5 mm thick)
with an individual photomultiplier readout is mounted in the front of each
crystal. It acts as "Veto" for charged
particles when photons or neutrons are detected in the TAPS. The 1.8 m
distance between the TAPS and the target is quite enough to separate pions
and protons using time of flight of the particle and its total energy
absorption in the TAPS.
Besides we will be able to identify neutrons and calculate
theirs energy by the time of flight. High granularity of the TAPS
provides a good tracking for both charge and neutral particles.

The described experimental setup is now available in the experimental hall
of the A2 collaboration at the continuous-wave electron accelerator
MAMI B \cite{mami1,mami2}. The Glasgow-Mainz photon tagging
facility \cite{tagg1,tagg2},
which provides photon beam for A2 collaboration experiments,
can tag bremsstrahlung photons in the energy range 40-820 MeV with
an intensity $\sim 0.6\times 10^6\gamma/$s in the lowest photon energy
tagger channel. The average energy resolution is 2~MeV.

\section{Background}

The charged and neutral particles CBTAPS detector combined
with the Glasgow-Mainz tagging facility is a very advantageous
instrument to search for and investigate the narrow six-quark states
in the pion photoproduction reactions. For this experimental setup
an optimal way to recognize the narrow six-quark states
is a reconstruction of invariant mass for three particles detected:
photon and two nucleons. To suppress the background
contribution to this spectrum it is necessary to additionally detect
the pion, neutral or charged, depending on the reaction channel.

The main anticipated background reactions are
\beq\label{dph}
&& \g+d\to\po+\pp+n+n, \nonumber \\
&& \g+d\to\po+\pn+p+p, \\
&& \g+d\to\po+\po+p+n' \nonumber
\eeq
and
\begin{equation}\label{rph}
\g+d\to\pi+\g+N+N .
\end{equation}
The double pion production reactions are important for the
photon energies more then 500 MeV and only in the case when one of
the photons from the $\po$ decay is not detected. But the $\po$ detection
efficiency for CBTAPS detector is over 85\% and it will reject most of
this background. Because we detect all secondary particles involved
we can apply a lot of different kinematic cuts which will further suppress
these processes.
The process (\ref{rph}) has the value of the total cross section similar
to the one under study. However,this background is distributed over full
mass region, whereas the SND photoproduction
gives contribution in a narrow region about the six-quark state mass.

\section{GEANT simulation}

To estimate expected yields for the SND formation we performed
Monte-Carlo simulation based on GEANT3 code \cite{geant}, in
which all relevant properties of the setup are taken into account.
Initial distributions for the event generator included differential
cross sections of the SND production calculated according to sec. 2.
The distribution of the SND decay probability (see Fig. \ref{width}) is
also taken into account. Besides the following beam conditions were used
\begin{itemize}
\item Incoming electron beam energy: \qquad\qquad$\,$ 880 MeV.
\item Tagged photon energy range: \qquad\quad 300 -- 820 MeV.
\item Maximal count rate in the tagger: \qquad$\,$ $6\times 10^5$ 1/s.
\item Tagging efficiency:\qquad\qquad\qquad\qquad\qquad\qquad$\;$ 50\%.
\end{itemize}

The results of the GEANT simulation of the invariant $\g NN$ mass
spectra for the production of the isovector SNDs with masses 1904, 1926,
1942, and 1980 MeV for the processes of the pion photoproduction
from the deuteron are presented in Fig. 14 for 300 hours of the beam time.
The productions of $D(1,1^-)$ in the reactions $\gp$ and $\gn$ are shown
in Fig. 14a and 14b, respectively. Fig. 14c shows the invariant $\g pn$
mass spectrum for the $D(1,1^+)$ in the process of $\pi^0$ meson
photoproduction from the deuteron. As seen from Fig 14, it is expected that
the SND can be easy extracted from the experimental data with the number
of standard deviations more than 10.

If SNDs really exist then, besides the peaks in $\g NN$ spectra, we must
also observe the corresponding peaks in $\g p$ and $\g n$ mass spectra
which connected with the dynamic of the SNDs decay. Such spectra, obtained
in the GEANT simulation, are represented in Fig. 15 for the process
$\g d\to\pi^0 D(1,1^+)$, $D(1,1^+)\to \g pn$ and in Fig. 16 for
$\g d\to\pi^- D(1,1^-)$, $D(1,1^-)\to \g pp$. Figs. 15a and 16a demonstrate
the spectra without an influence of the detectors and Figs. 15b and 16b
with an influence of the detectors. The observation of the peaks in
the $\g N$ mass spectra is the necessary condition to prove the SND
existence.

The spectra in Figs. 15a and 16a were obtained assuming that SNDs decay
without of the $N^*$ production. As a result, the widths of the peaks in
these figures are about a few MeV. If the exotic baryons $N^*$ exist, their
decay widths have to be $\ll 1$ MeV. Therefore, experiments with the
mass resolution $\lesssim 1$ MeV  can give a conclusion about the existence
of the exotic baryons $N^*$ with the small masses.

The expected yields of the SNDs as a result of the simulation of the
isovector SND production in the processes of the pion photoproduction
from the deuteron at the Mainz microtron MAMI in the photon energy region
300--820 MeV for the beam time of 300 hours are listed in table 2 for
the different masses of SNDs. The expected mass resolution $\sigma_M$ are
presented here too. As seen from this table the biggest yield of
$D(1,1^-)$ is expected in the process of the $\pi^-$ meson photoproduction.
The photoproduction of $\pi^0$ meson results mainly in the $D(1,1^+)$
formation. So investigation of the considered in this work processes
give possibility both to observe the SNDs and determine their quantum numbers.

\begin{table}
\centering
\begin{tabular}{|c|c|c|c|c|c|c|} \hline
$M$& \multicolumn{2}{|c|}{$\g d\to\pp D(1,1^-)$}&
\multicolumn{2}{|c|}{$\g d\to\pn D(1,1^-)$}&
\multicolumn{2}{|c|}{$\g d\to\po D(1,1^+)$} \\ \cline{2-7}
(MeV)&  &  &  &  &  &  \\
     &yields&$\sigma_M$
     &yields&$\sigma_M$
     &yields&$\sigma_M$       \\ \hline
     &  &  &  &  &  &     \\
1904 &380&4.0&4720&4.0&1300&3.8 \\ \hline
     &  &  &  &  &  &     \\
1926 &460&4.6&5270&4.6&1350&4.4 \\ \hline
     &  &  &  &  &  &     \\
1942 &480&5.7&5120&5.5&1340&5.6 \\ \hline
     &  &  &  &  &  &     \\
1980 &560&6.9&4830&7.0&1180&6.9 \\ \hline
\end{tabular}
\caption{ Expected yields of the SNDs
and the mass resolutions \mbox{($\sigma_M$ (MeV))}}
\end{table}


\section{Summary}

\begin{itemize}
\item A search for narrow six-quark states in the reactions
$\gp$, $\gn$, and $\go$ at \mbox{MAMI-B} is proposed.
\item The masses of the SNDs will be reconstructed by a measurement
of the photon and two nucleon.
\item Using of the deuteron target allows avoiding uncertainties taken place
in the experiment at INR.
\item Using of the Crystal Ball spectrometer and TAPS allows one to
detect $\g$, $p$ and $n$ with good accuracy and
suppress essentially the background.
\item This experiment will give possibility to observe the SNDs in the
mass region from 1880 up to 2000 MeV with good enough precision.
\item A comparison of the results obtained for the reactions under study
will allow the quantum numbers of the SNDs ($T$, $J^P$)
to be determined.
\item Study of the $\g p$ and $\g n$ mass spectra will give
an additional information about the nature of the observed dibaryon states
and a possibility of existence of exotic baryons with small masses.
\end{itemize}

\newpage

\newpage

\newpage

\newpage

\newpage
\begin{figure}[ht]
\centerline{
\epsfxsize=16cm
\epsfysize=20cm
\epsffile{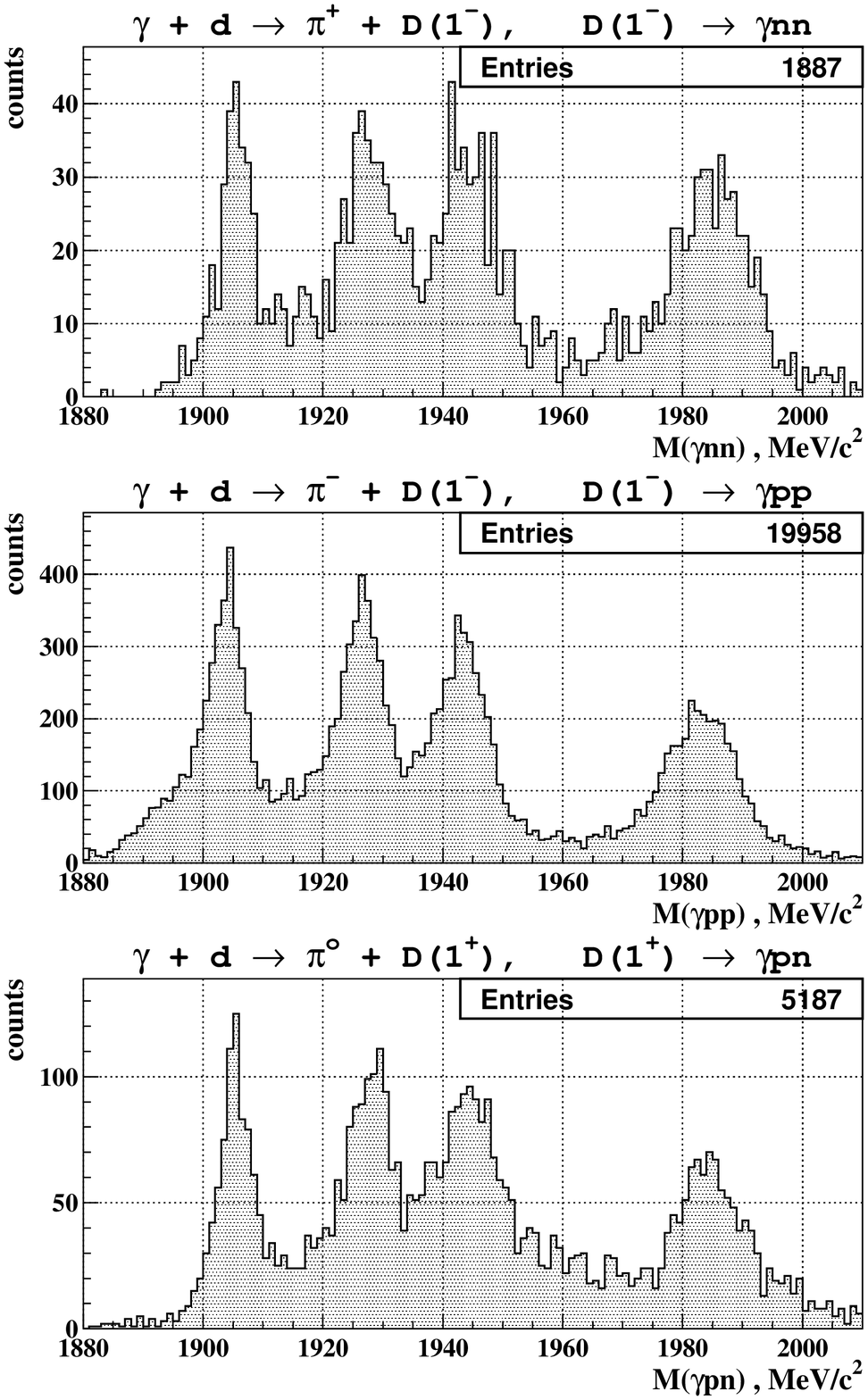}}
\caption{GEANT simulation of the SNDs with $M=1904$, 1926,
1942, and 1980 MeV production in the processes $\gp$, $\gn$,
and $\go$ for 300 hours of beam time.}
\label{all}
\end{figure}

\newpage
\begin{figure}[ht]
\centerline{
\epsfxsize=16cm
\epsfysize=20cm
\epsffile{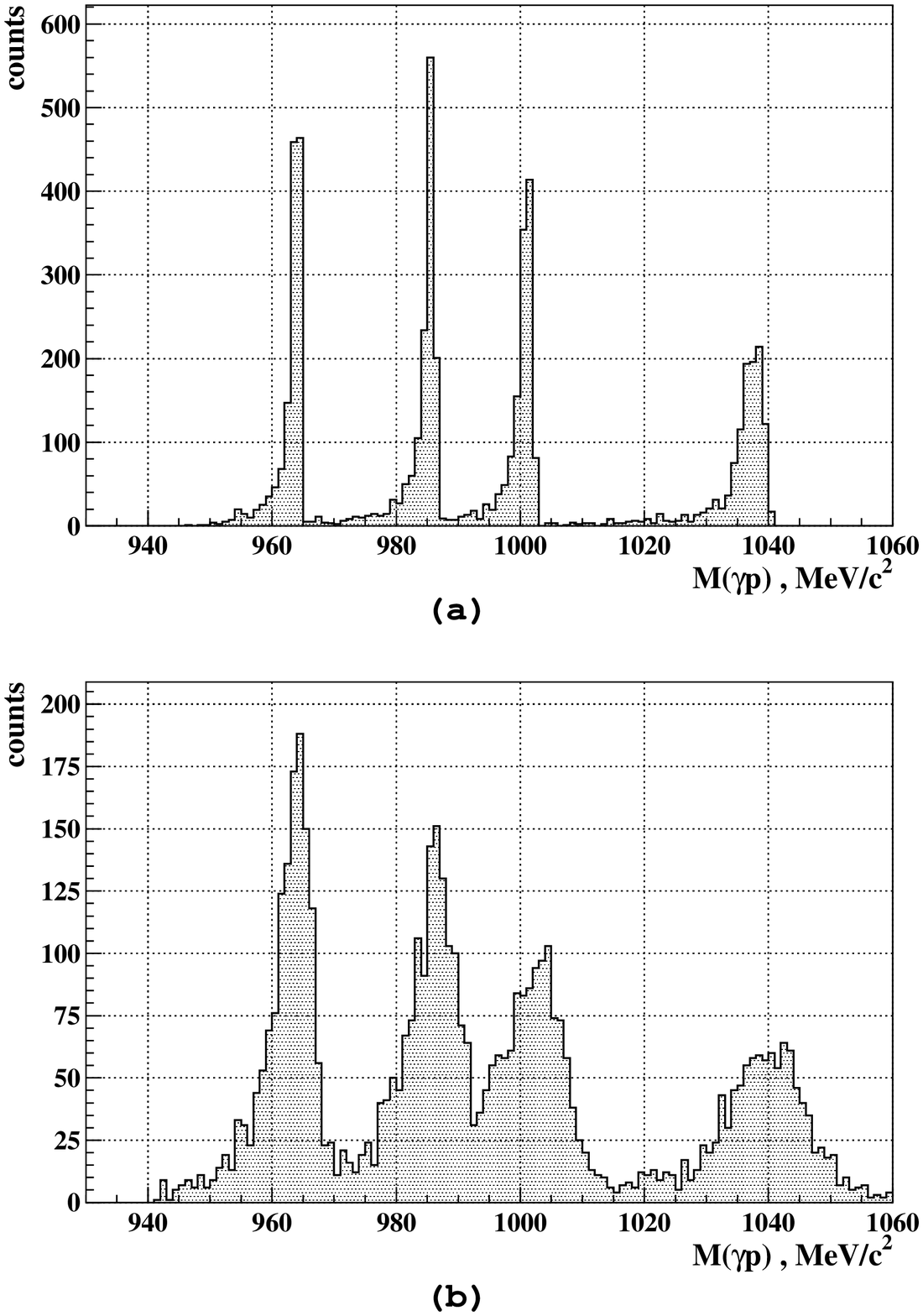}}     
\caption{GEANT simulation of the invariant $\g p$ and $\g n$
mass spectra for the reaction $\go$; a -- without an influence of
the detectors; b -- with an influence of the detectors}
\label{pn}
\end{figure}

\newpage
\begin{figure}[ht]
\centerline{
\epsfxsize=16cm
\epsfysize=20cm
\epsffile{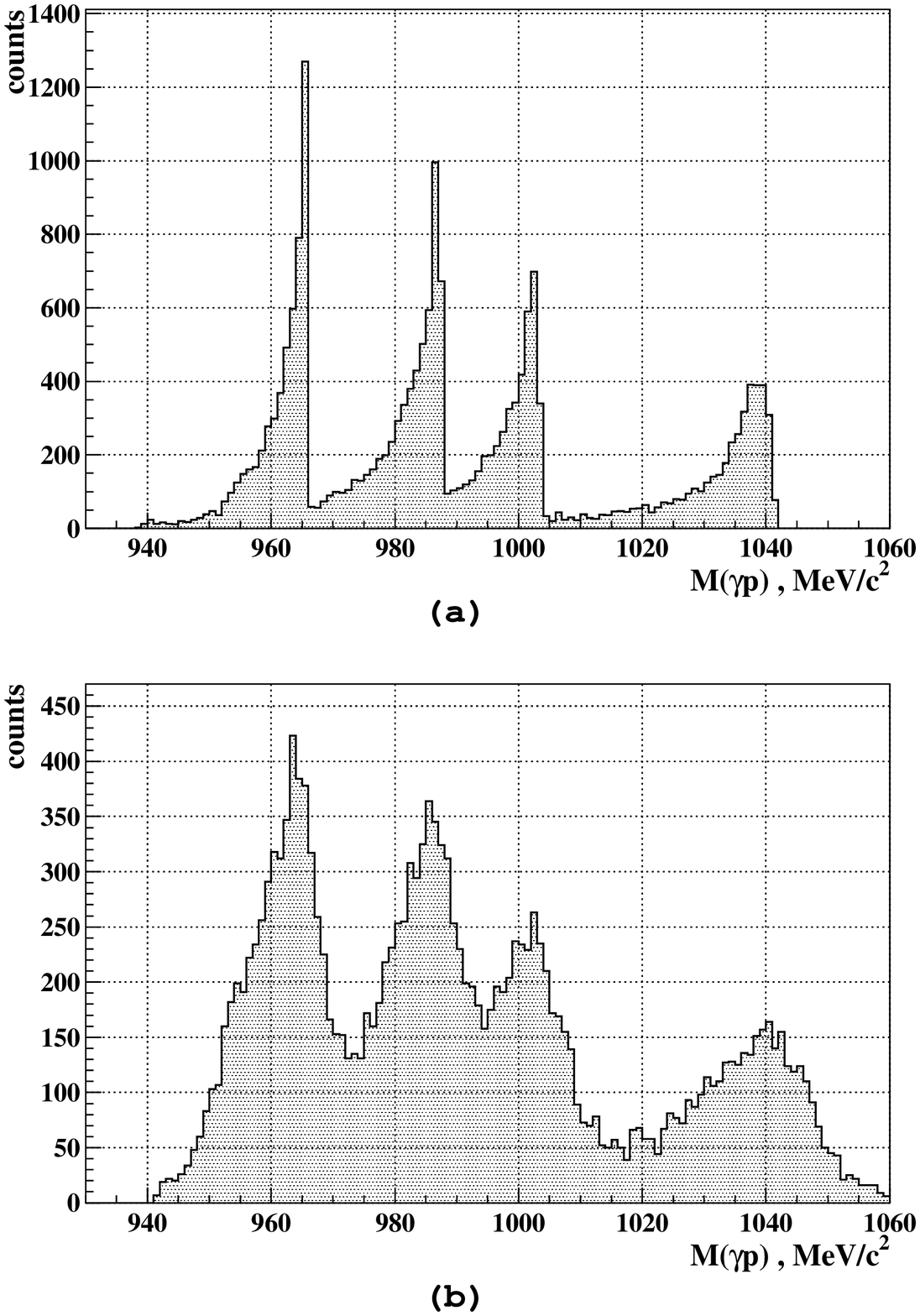}}    
\caption{GEANT simulation of the invariant $\g p$
mass spectra for the reaction $\gn$; a-- without an influence of
the detectors; b -- with an influence of the detectors}
\label{pp}
\end{figure}

\end{document}